\documentclass[preprint, aps, prf, floatfix, longbibliography, groupedaddress, showkeys]{revtex4-1}

\usepackage{tabularx}
\usepackage{graphicx}
\usepackage{natbib}
\usepackage{pifont}
\usepackage{soul}
\usepackage{amsmath}
\usepackage{amssymb}
\usepackage{color}
\usepackage{mathtools}
\usepackage{relsize}
\usepackage{wasysym}
\usepackage[hidelinks=true]{hyperref}
\usepackage{bm}                 

\newcommand{\vol}{\mathop{\ooalign{\hfil$V$\hfil\cr\kern0.08em--\hfil\cr}}\nolimits}

\newcommand{\bs}[1]{\boldsymbol{#1}}

\newcommand{\norm}[1]{\left\lVert#1\right\rVert}

\newcommand{\mrm}[1]{\mathrm{#1}}

\begin{document}


\title{A Koopman-based framework for forecasting the spatiotemporal evolution of chaotic dynamics with nonlinearities modeled as exogenous forcings}


\author{M. A. Khodkar$^1$}\thanks{mkhodkar@rice.edu}
\author{Pedram Hassanzadeh$^{1,2}$}\thanks{pedram@rice.edu}
\author{Athanasios Antoulas$^{3, 4, 5}$}

\affiliation{$^1$Department of Mechanical Engineering, Rice University, Houston, TX 77005, USA \\
$^2$Department of Earth, Environmental and Planetary Sciences, Rice University, Houston, TX 77005, USA \\
$^3$Department of Electrical and Computer Engineering, Rice University, Houston, TX 77005, USA \\
$^4$Baylor College of Medicine, 1 Baylor Plaza, Houston, TX 77030, USA  \\
$^5$Data-Driven System Reduction and Identification (DRI) Group, Max Planck Institute for Dynamics of Complex Technical Systems, Sandtorstrasse 1, 39106 Magdeburg, Germany}

\date{\today}

\begin{abstract}

We introduce a data-driven method and shows its skills for spatiotemporal prediction of high-dimensional chaotic dynamics and turbulence. The method is based on a finite-dimensional approximation of the Koopman operator where the observables are vector-valued and delay-embedded, and the nonlinearities are treated as external forcings. The predictive capabilities of the method are demonstrated for well-known prototypes of chaos such as the Kuramoto-Sivashinsky equation and Lorenz-96 system, for which the data-driven predictions are accurate for several Lyapunov timescales. Similar performance is seen for two-dimensional lid-driven cavity flows at high Reynolds numbers.  

\end{abstract}

\maketitle


\section{Introduction \label{Section:Intro}}

Predicting the spatiotemporal evolution of high-dimensional and nonlinear dynamical systems, such as turbulent flows, has been of long-standing interest in science and engineering \citep{Lorenz1963,Box2015}. For example, forecasting turbulent flows plays a key role in controlling and optimizing various engineering systems (e.g. wind farms) and predicting the state of the atmosphere and/or ocean (e.g. day-to-day weather) \citep{van2016long,duriez2017machine,majda2012challenges}. For many of these problems, an objective of particular interest and wide-ranging applications is predicting extreme events at some useful lead time \citep{bauer2015quiet,majda2018model,farazmand2017variational}.  

Data-driven prediction of chaotic dynamics and turbulent flows has received significant attention in recent years, in particular, for problems in which the high-dimensional, nonlinear governing equations cannot be solved fast enough to be useful (e.g. for online control/optimization), or in which some of the physical processes (and thus the governing equations) are not fully understood but observational data from the past are available (e.g. the weather/climate systems) \citep{wunsch1999interpretation,van2007empirical,cavanaugh2015skill,Giannakis2017,comeau2017data,Khodkar2018}. Rapid advances have been recently made in this area based on using techniques from machine learning or approximating the Koopman operator. These approaches involve using the past data to build/train a model that can produce accurate and fast predictions about the future spatiotemporal evolution of the flow. Promising results for prototypes of chaotic dynamics, e.g. Lorenz-63/96 and Kuramoto-Sivashinsky (K-S) equations, or classical fluid examples such as vortex shedding past a cylinder and homogenous isotropic turbulence, have been reported using machine learning methods such as long short-term memory (LSTM) networks, physics-informed neural networks, an reservoir computing \citep{McDermott2017, yu2017long, Vlachas2018, Pathak2018, Raissi2019, Mohan2019, McDermott2019, Ashesh2019}.       

The Koopman operator \citep{koopman1931hamiltonian}, which is an infinite-dimensional linear operator that shifts observables forward in time, offers a powerful framework for analyzing and tackling nonlinear systems such as fluid flows \citep{mezic2005spectral,mezic2013analysis}. Data-driven, finite-dimensional approximations of the Koopman operator, using methods such as dynamic mode decomposition (DMD) and its variants \citep{Schmid2010, rowley2009spectral,Tu2014,williams2015data,Arbabi2017,Arbabi2017study,korda2018convergence}, have been extensively used to analyze various flows in recent years \citep{mezic2013analysis,Rowley2017}. Coupling the Exact DMD of \citet{Tu2014} with the delay-embedding theorem of \citet{Takens1981}, \citet{Arbabi2017} introduced the Hankel-DMD method and proved its convergence to the Koopman operator. \citet{Korda2018} have utilized this method for the short-term forecasting of a forced Van der Pol oscillator as well as for the feedback control of a bilinear DC motor. This concept was further employed by \citet{Arbabi2018} to develop a predictive-control framework for the purpose of stabilizing a two-dimensional (2D) lid-driven cavity flow that has bifurcated to a limit cycle, with emphasis on the delay-embedding of measurements to reduce the number of placed sensors. 

Furthermore, Giannakis and collaborators have shown that projecting the delay-embedded data onto the eigenfunctions of Laplace-Beltrami \citep{Giannakis2012} or Koopman operator \citep{Giannakis2017} leads to the frameworks which are skillful in model reduction, mode decomposition and forecasting of time series for ergodic dynamical systems with strongly nonlinear phenomena such as intermittitencies or energy bursts. \citet{Brunton2017} have built another successful data-driven model, skilled in detecting the low-probability events of chaotic dynamics, based on the sparse identification of nonlinear dynamics (SINDy) described in \citet{Brunton2016b}, and the representation of chaos as an intermittent forcing in the form of the principal component (PC) of the last retained mode given by the singular value decomposition (SVD) of the delay-embedded data. They notice that the statistics of this intermittent forcing is non-Gaussian, and it is connected to the PCs of other retained modes present in the state vector in a nonlinear fashion. The model, referred to as Hankel alternative view of Koopman (HAVOK) by the authors, then accurately predicts the lobe-switching events of various chaotic attractors such as Lorenz-63, R\"{o}ssler and double-pendulum, one time unit ahead of their occurrences. 

The corresponence between the aforementioned Koopman-based methods which benefit from delay-embedded measurements and the system identification method matrix pencil (which is called Loewner method when used in frequency-domain) in time-domain is mathematically established by \citet{Thanos2012}. This approach has been recently used for data-driven model reduction of nonlinear systems such as Burgers' equation \citep{Thanos2016, Thanos2019a, Thanos2019b}. The underlying connections between the two methods have been further investigated in \citet{Pogorelyuk2018}.

Despite the success of previous models, data-driven spatiotemporal prediction of high-dimensional and highly chaotic systems for reasonably long times is still the subject of ongoing research. Towards this end, we develop a data-driven Koopman-based method which models the nonlinearities as external forcings (actuations), while the observables forming the state vector are vector-valued, delay-embedded and linear. The unknown maps (matrices) appearing in the method are found using the method of \citet{Brunton2016a}, known as dynamic mode decomposition with control (DMDc). The paper is organized as follows. The mathematical derivation of the method and its implementation for some well-known prototypes of chaos are discussed in Sec.~\ref{Section:Theory}. Further results regarding the predictive capabilities of the method for chaotic dynamical systems (Lorenz-63, K-S and Lorenz-96) as well as a fluid test case (a 2D lid-driven cavity flow at high Reynolds numbers) are presented in Sec.~\ref{Section:Results}. Section~\ref{Section:Conclusion} summarizes the methodology and the findings, and outlines the prospects of the proposed method.

\color{black}

\section{Methodology and applications to some chaotic test cases \label{Section:Theory}}

In the following, we present a Koopman-based data-driven method, which enables the spatiotemporal prediction of chaotic dynamics such as the K-S equation and Lorenz-96 system. We also shed some insights into the accurate representation of nonlinearity. General guides regarding the proper selection of the method's parameters are also provided in Appendix \ref{Section:Parameter}.  

\subsection{A data-driven predictive framework for the Kuramoto-Sivashinsky equation \label{Section:Model}}

We use the K-S equation, a widely-used prototype for spatiotemporal chaotic systems, as an example to formulate our proposed data-driven method. The K-S equation is described by
\begin{eqnarray}
	\frac{\partial u}{\partial t} = -u \frac{\partial u}{\partial x} - \frac{\partial^2 u}{\partial x^2} - \frac{\partial^4 u}{\partial x^4} + \eta \cos(2\pi x/\delta) \, , \label{eqn:KS}  
\end{eqnarray}
where $u$ denotes the K-S variable. The last term in Eq. (\ref{eqn:KS}) is a periodic forcing which causes spatial inhomogeneity \citep{Pathak2018}. Here, we take $\delta = L/2$, where $L$ is the domain length. The choice for the number of collocation points $n$ depends on the system's chaoticity, which is increased by the domain length $L$, so $n$ will be reported individually for each case. A periodic boundary condition is enforced, and a pseudospectral solver with the classic fourth-order Runge-Kutta (RK4) is used for integrating (\ref{eqn:KS}), to construct training sets with $N = 70000$ data points and sampling interval $\tau = 0.02\tau_d$, where $\tau_d$ is the decorrelation timescale of the principal component of the leading POD mode (PC1). The training set refers to the part of the dataset used for building the Koopman-based predictive method, which is fully separate from the dataset used for examining the performance of the model, called the testing set. The length of our training sets and sampling intervals are identical to those of \citet{Pathak2018}. We have also constructed $20$ independent testing sets, each with $30000$ samples, to evaluate the performance of the proposed data-driven methods. 

Relative errors are calculated as 
\begin{equation}
    E(t) = \norm{\bs{u}_{pred} - \bs{u}_{num}}_2 /\norm{\bs{u}_{num}}_2 \, , \label{eqn:Error}  
\end{equation}
and are averaged over all these testing sets. Note that $\norm{\,}_2$ denotes the Euclidean norm, and $\bs{u}_{pred}$ and $\bs{u}_{num}$ represent the values given by the data-driven predictive methods and the pseudospecral numerical solver, respectively, while the latter is taken as ground truth. Another measure of error can also be defined as $E_{ave} = 1/t_l \int_0^{t_l} E(t) \mrm{d}t$, which is simply the temporal mean of $E(t)$ from the starting point of prediction ($t = 0$) to the divergence time $t_l$, where $t_l$ corresponds to the time at which $E$ first exceeds $0.3$.

Suppose $N$ samples of vector-valued observables $\bs{u}^i$ obtained from running the numerical solver are arranged in the Hankel matrix $\mathcal{H}$     
\begin{eqnarray}
    \mathcal{H} = \left[
   \begin{array}{cccc}
   \bs{u}^1  & \bs{u}^2     & \dots & \bs{u}^{N-q+1}  \\
   \bs{u}^2  & \bs{u}^3     & \dots & \bs{u}^{N-q+2}  \\
   \vdots    & \vdots       & \dots & \vdots          \\
   \bs{u}^q  & \bs{u}^{q+1} & \dots & \bs{u}^{N} 
   \end{array}  
\right]  \, ,  \label{eqn:HankelMatrix}
\end{eqnarray}
where $\bs{u}^i \in \mathbb{R}^n$ is sampled at $t = i\tau$, and $q$ is the delay-embedding dimension. The size of $\mathcal{H}$ is then $(n \times q) \times (N-q+1)$. Following the Hankel-DMD formulation of \citet{Arbabi2017}, we construct 
\begin{eqnarray}
\mrm{X} = \mathcal{H}(:, 1:N-q) \, , \quad \mrm{Y} = \mathcal{H}(:, 2:N-q+1) \, , \label{eqn:LaggedMatrix}  
\end{eqnarray}
and conduct reduced SVD to obtain $\mrm{X} = \mrm{U}\mrm{S}\mrm{V}^*$, in the subspace of leading $r$ singular vectors ($*$ indicates conjugate transpose). The data-driven approximation of Koopman operator using Hankel-DMD is computed as \citep{Arbabi2017}
\begin{eqnarray}
\mrm{A}_{\mrm{HDMD}} = \mrm{U}^* \mrm{YV} \mrm{S}^{-1},  \,  \label{eqn:DMD}
\end{eqnarray}
which has the size $r \times r$.  

Once $\mrm{A}_{\mrm{HDMD}}$ is calculated, a future vector-valued observable $\bs{u}^{m+1}$, that was not part of the training set, can be predicted from
\begin{eqnarray}
	\bs{\mathcal{U}}^{m+1, r} = \mrm{A}_{\mrm{HDMD}} \ \bs{\mathcal{U}}^{m, r}  \, , \label{eqn:Pred}
\end{eqnarray}
where $\bs{\mathcal{U}}^{m+1, r}$, a vector of length $r$, is $\bs{\mathcal{U}}^{m+1} = [\bs{u}^{m-q+2} \ \bs{u}^{m-q+3} \ \cdots \ \bs{u}^{m+1}]^T$ ($T$ indicates transpose), a vector of length $n \times q$, projected onto the subspace of first $r$ singular vectors. The first block of $\bs{\mathcal{U}}^{m+1}$ is then taken as the prediction of the new state. All values in $\bs{\mathcal{U}}^{m} = [\bs{u}^{m-q+1} \ \bs{u}^{m-q+2} \ \cdots \ \bs{u}^{m}]^T$ and its projection onto the subspace of retained singular vectors, $\bs{\mathcal{U}}^{m, r}$, are either known from the initial condition or already predicted. We notice that the first $q-1$ predictions are the reconstruction of initial condition, and only the $q^{\mrm{th}}$ prediction is the forecast value of the future state. Hereafter, we refer to this method, which is thoroughly based on the Hankel-DMD method introduced by \citet{Arbabi2017}, as M1.  

One may suspect that the linear combination of a finite number of DMD modes, specifically those given by a fairly short dataset of linear observables, cannot accurately reproduce the nonlinear characteristics of a chaotic dynamics for a reasonably long period of time. Prior to investigating this hypothesis and the performance of M1, we attempt to develop a modified method (M2), which is specialized to tackle the issue of the nonlinearity. Inspired by HAVOK model of \citet{Brunton2017}, in which adding a forcing term to the linear model is seen to approximate the nonlinear dynamics more accurately and yield better predictions, we incorporate the nonlinear effects in the form of external forcings, so that any dynamical system can be modeled as
\begin{eqnarray}
	\bs{u}^{m+1} = \mrm{A} \bs{u}^{m} + \mrm{B} \bs{f}^m \, . \label{eqn:M2}
\end{eqnarray} 
Unlike \citep{Brunton2017} which uses the last retained singular vector as forcing, we choose the forcing term in a physics-driven fashion. For instance, when some knowledge of the governing equations is available, or one can intuitively speculate the form of nonlinearity, the forcing term can be chosen according to that knowledge or intuition. Consequently, for the K-S equation, forcing vector $\bs{f}^i$ includes the square of $u$ at the same snapshot and every grid point, i.e. $\bs{f}^i = \big[(u^i_1)^2 \ (u^i_2)^2 \ \cdots \ (u^i_n)^2 \big]^T$. The delay-embedded form of Eq.~(\ref{eqn:M2}) reads
\begin{eqnarray}
	\bs{\mathcal{U}}^{m+1} = \mrm{A} \bs{\mathcal{U}}^{m} + \mrm{B} {\bs{\mathcal{F}}^m} \, , \label{eqn:M2_delay}
\end{eqnarray}
where the definition of $\bs{\mathcal{U}}^m$ is the same as before. Now the forcing vectors should also be sampled at each snapshot and sorted in the following Hankel matrix $\mathcal{F}$ 
\begin{eqnarray}
    \mathcal{F} = \left[
   \begin{array}{cccc}
   \bs{f}^1  & \bs{f}^2     & \dots & \bs{f}^{N-q}  \\
   \bs{f}^2  & \bs{f}^3     & \dots & \bs{f}^{N-q+1}  \\
   \vdots    & \vdots             & \dots & \vdots          \\
   \bs{f}^q  & \bs{f}^{q+1} & \dots & \bs{f}^{N-1} 
   \end{array}  
\right]  \, .  \label{eqn:HankelMatrix_forcing}
\end{eqnarray}
Note that however for the K-S system the state vector $\bs{u}$ and the forcing term $\bs{f}$ are of the same size, but in general the forcing vector could be much larger and of the length $n' \gg n$, depending on the form of nonlinearities and the number of nonlinear processes in the dynamical system. The size of $\mathcal{F}$ is thus $(n' \times q) \times (N-q)$.

\begin{figure}
  \centerline{\includegraphics[width=0.70\textwidth]{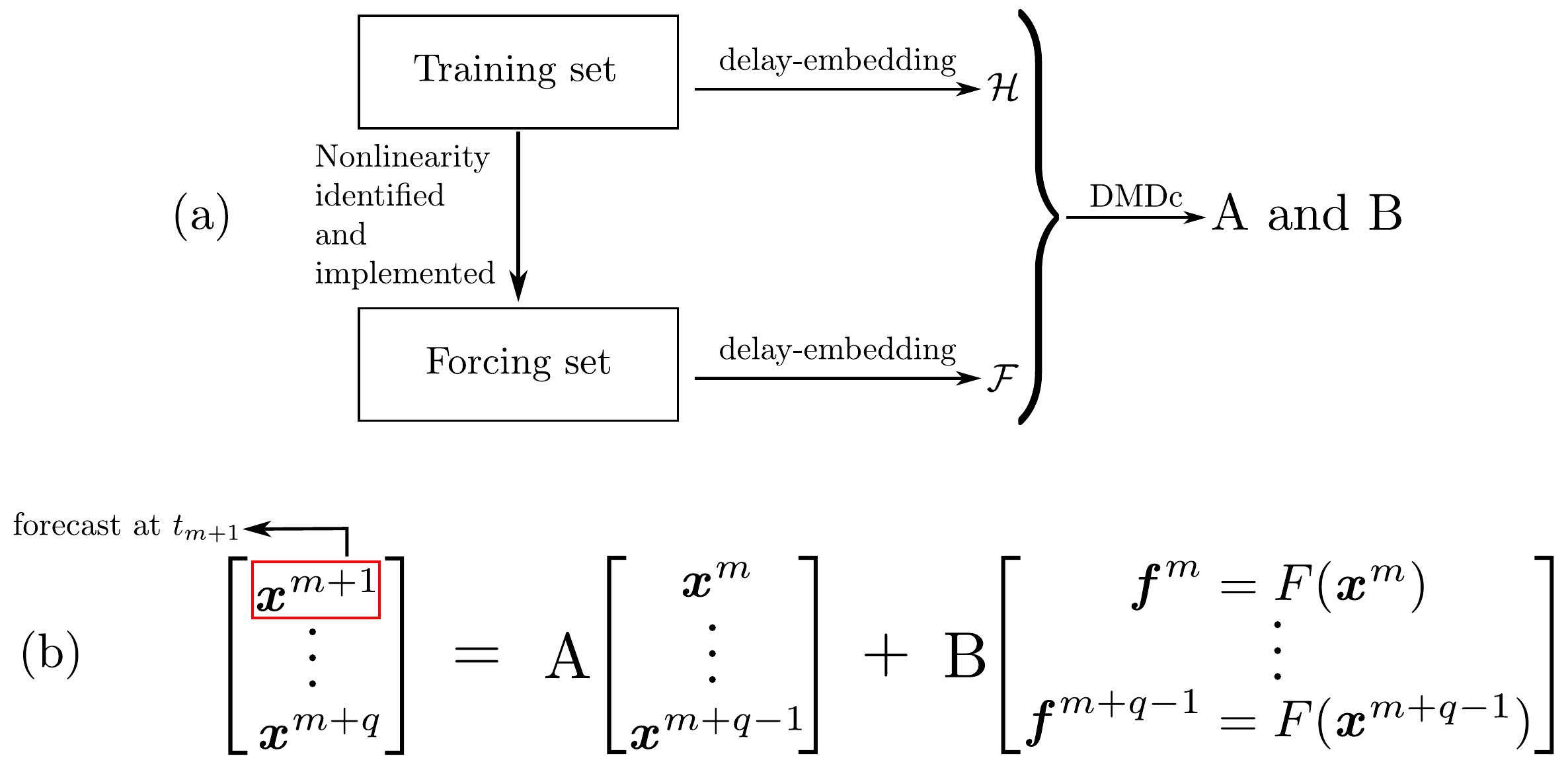}}
 \caption{Schematic of data-driven method M2: (a) Training on vector-valued and time-delay-embedded observables yields unknown matrices $\mrm{A}$ and $\mrm{B}$, while nonlinearites are modeled as external forcings and DMDc of \citet{Brunton2016a} is utilized. (b) Matrices $\mrm{A}$ and $\mrm{B}$ can then be employed for the spatiotemporal forecasting of the dynamical system. All vectors on the right-hand side are either known from initial condition or already predicted. Note that $\bs{f}^i$ is a function of $\bs{x}^i$.}
\label{Fig:Cartoon}
\end{figure}

The unknown maps $\mrm{A}$ and $\mrm{B}$ are then found using DMDc method presented in \citet{Brunton2016a}, which simply minimizes the Frobenius norm $\norm{Y - \mrm{A}X - \mrm{B}\mathcal{F}}_F$ to achieve
\begin{eqnarray}
	\mrm{A} &=& \hat{\mrm{U}}^* \mrm{Y} \tilde{\mrm{V}} \tilde{\mrm{S}}^{-1} \tilde{\mrm{U}}_1^*\hat{U}   \, , \nonumber \\
	\mrm{B} &=& \hat{\mrm{U}}^* \mrm{Y} \tilde{\mrm{V}} \tilde{\mrm{S}}^{-1} \tilde{\mrm{U}}_2^*          \, . \label{eqn:map}
\end{eqnarray}
Here, $\mrm{Y} = \hat{\mrm{U}}\hat{\mrm{S}}\hat{\mrm{V}}^{*}$, while the truncation value is taken as $r$, i.e. $\hat{\mrm{U}} \in \mathbb{R}^{nq \times r}$, $\hat{\mrm{S}} \in \mathbb{R}^{r \times r}$ and $\hat{\mrm{V}} \in \mathbb{R}^{(N-q)\times r}$, and $[\mrm{X} \ \mathcal{F}]^T = \tilde{\mrm{U}}\tilde{\mrm{S}}\tilde{\mrm{V}}^{*}$ with the truncation value selected as $p$, so that $\tilde{U} \in \mathbb{R}^{(n+n')q \times p}$, $\tilde{\mrm{S}} \in \mathbb{R}^{p \times p}$ and $\tilde{\mrm{V}} \in \mathbb{R}^{(N-q)\times p}$. $\tilde{U_1}$ and $\tilde{U_2}$ are made up of the first $nq$ and the remaining $n'q$ rows of $\tilde{\mrm{U}}$, respectively. Note that $\mrm{A}$ and $\mrm{B}$ are calculated in a reduced-dimension subspace, and they have the respective sizes $r \times r$ and $r \times nq$. Similar to M1, once training is done and the unknown maps are calculated, the first $q$ data points in the testing set will be used to initialize the state vector, i.e. these points are not predicted. \textit{Nonetheless all results shown after $t = 0$ are newly predicted values by the data-driven methods, and were not part of the initial condition.} Figure \ref{Fig:Cartoon} summarizes the training and forecasting steps of this method.  

\begin{figure}
  \centerline{\includegraphics[width=1.0\textwidth]{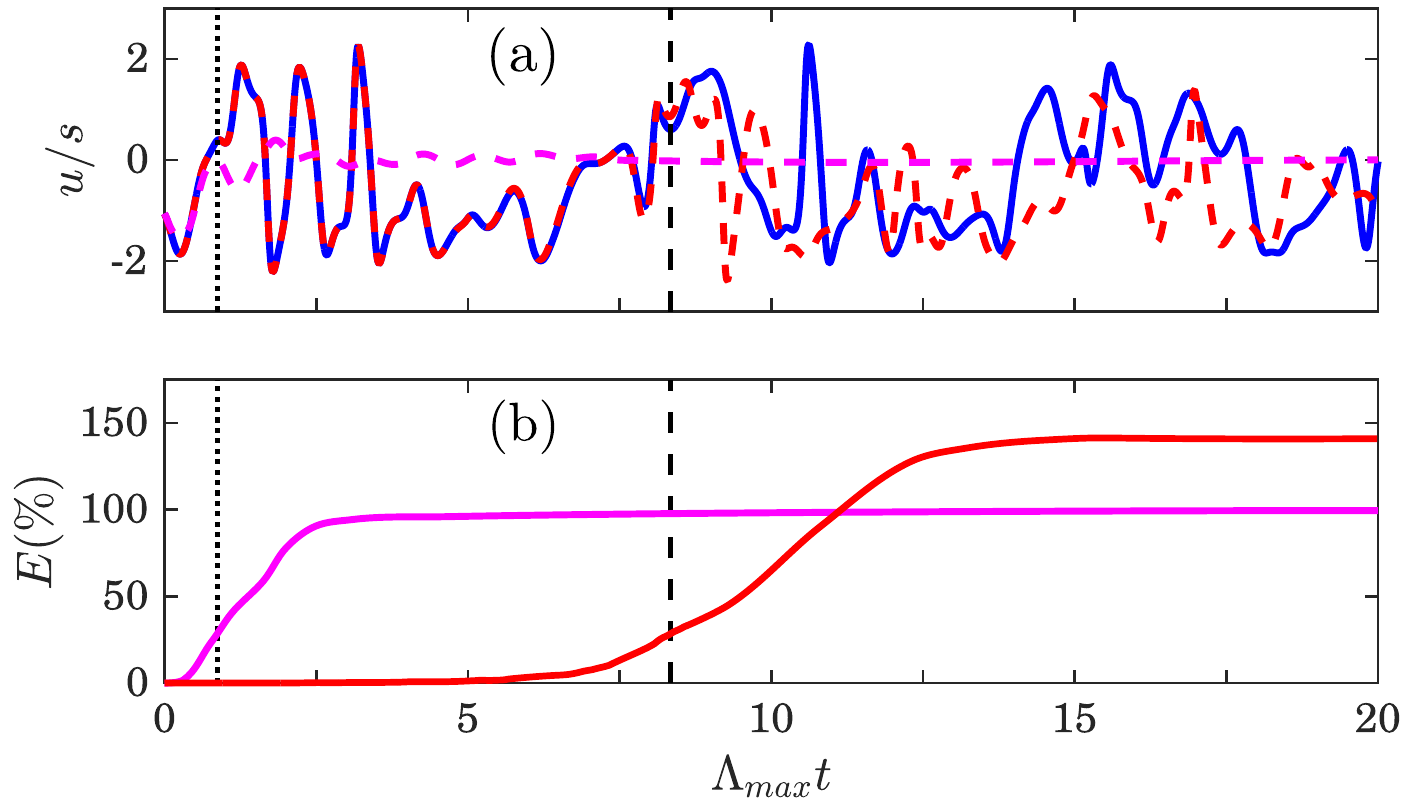}}
 \caption{(a) Predictions given by M1 (dashed magenta) and M2 (dashed red) for the time series of $u$ at $x = 8$, compared to the ground truth obtained via the numerical integration of Eq. (\ref{eqn:KS}) (solid blue). Note that $y$-axis is normalized by the standard deviation $s$ of testing data. (b) Variation of relative error $E$ rendered by each method with time scaled by Lyapunov timescale $1/\Lambda_{max}$. Again, magenta and red lines represent the results of M1 and M2, respectively, while the vertical dotted and dashed lines mark the time at which the prediction of each method diverges from the actual data. The studied K-S system is unforced with the domain length $L = 22$, the number of collocation points $n = 64$, and the attractor dimension $D_{KY} = 5.20$, computed based on the Kaplan-Yorke formulation \citep{Kaplan1978}.}
\label{Fig:KS_TS}
\end{figure}

\begin{figure}
  \centerline{\includegraphics[width=0.95\textwidth]{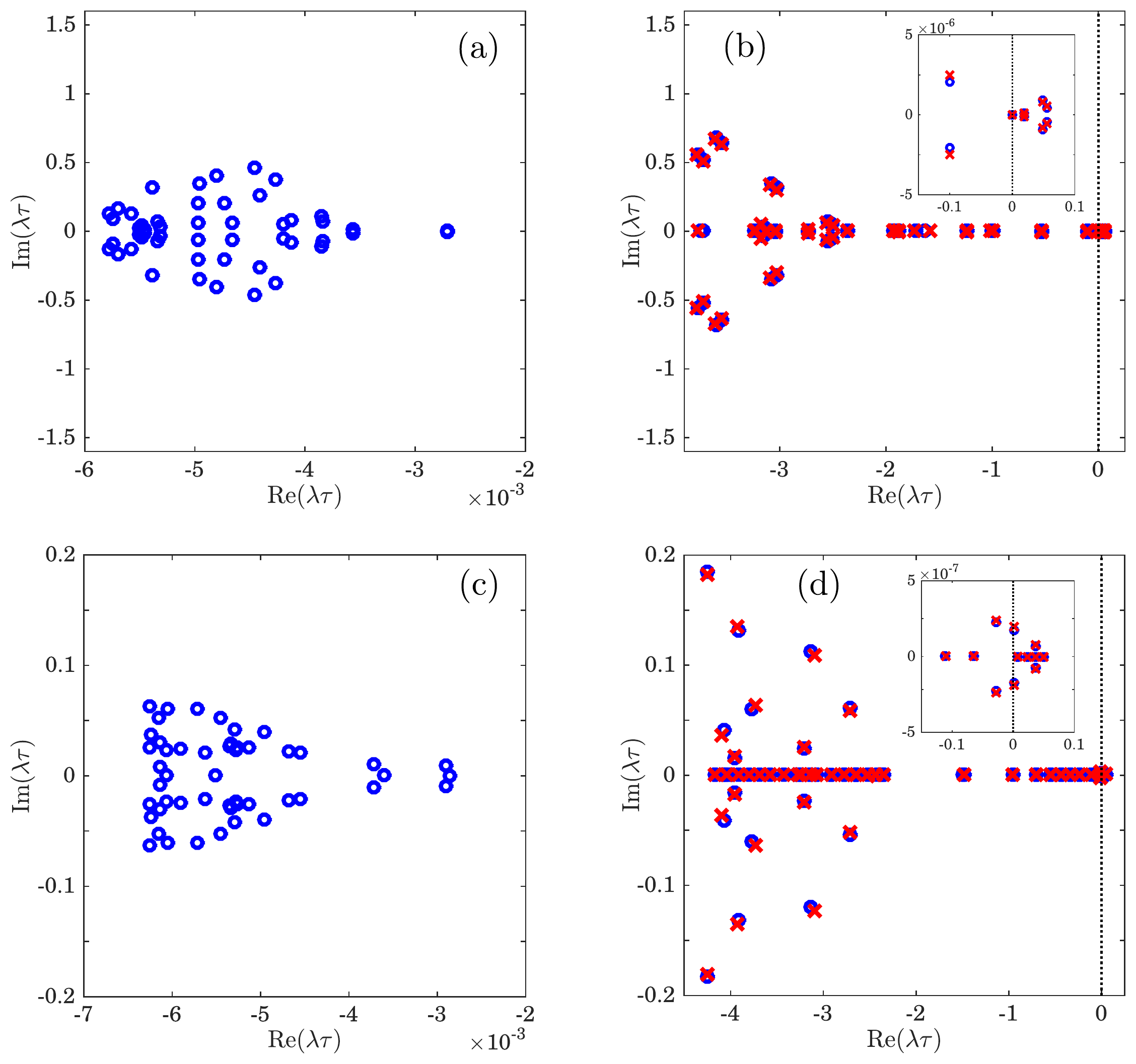}}
 \caption{Eigenvalues $\lambda$ of $\exp(\tau \mrm{A})$ for K-S equations with $L = 22$ (top) and $L = 100$ (bottom), normalized by sampling frequency $1/\tau$, and calculated via (a) M1 with $(r, q) = (2500, 50)$, (b) M2 with $(r, p, q) = (400, 400, 20)$, (c) M1 with $(r, q) = (2200, 40)$, and (d) M2 with $(r, p, q) = (800, 800, 10)$. The blue circles show the eigenvalues provided by the entire training set with $70000$ data points, while the red crosses in panels (b) and (d) indicate the eigenvalues identified when half the training set is used. The close agreement between the two suggests that the eigenvalues of the systems are captured robustly. The vertical dotted lines in these panels mark the imaginary axis. For clarity, only the first $50$ eigenvalues are depicted, and the smaller panels within panels (b) and (d) focusing on the eigenvalues of M2 with growing modes are also included. We notice that except for the few leading eigenvalues, modes given by M2 decay at much faster rates than those given by M1. The results of the case with $L = 100$ will be further discussed in Sec.~\ref{section:KS}.}
\label{Fig:EV_KS}
\end{figure}

Figure \ref{Fig:KS_TS}(a) displays the temporal evolution of $u (x = 8, \ t)$ predicted by M1 (dashed magenta) and M2 (dashed red), and compares them to the actual data (truth) obtained via the numerical integration of Eq. (\ref{eqn:KS}). \textit{We reiterate that all results shown for M1 and M2 after $t = 0$ are new forecast values, and were not used during training and building the model or as a part of the initial condition.} Time is scaled by Lyapunov timescale $1/\Lambda_{max}$, where the leading Lyapunov exponent $\Lambda_{max}$ is calculated following \citet{Wolf1985}. As shown in this figure, predictions rendered by the conventional Hankel-DMD method (M1) diverge from the testing data fairly rapidly in less than a Lyapunov timescale, and after that, predictions gradually decay to zero. This can be attributed to the fact all eigenvalues of $\mrm{A}_{\mrm{HDMD}}$ fall inside the unit circle, with most of them located in its vicinity, meaning all the modes corresponding to these eigenvalues are decaying (Fig.~\ref{Fig:EV_KS}(a) and (c)). To distinguish between the eigenvalues more clearly, Fig.~\ref{Fig:EV_KS} shows the eigenvalues $\lambda$ of $\exp(\tau \mrm{A})$. Therefore, the eigenvalues of $\mrm{A}$ inside/outside the unit circle correspond to the eigenvalues of $\exp(\tau \mrm{A})$ to the left/right of the imaginary axis. It is noteworthy that the special case of $q = 1$, leading to the conventional Exact DMD, does not reveal any predictive skill, so that its predictions become inaccurate in less than ten iterations, or $0.1/\Lambda_{max}$. On the other hand, the inclusion of nonlinearity in the form of external forcings has substantially improved the performance of the predictive framework so that, compared to the best results of M1, prediction horizon $t_l$ is increased by a factor larger than $10$. However, usually a few of eigenvalues of $\mrm{A}$ given by M2 (six in the case with $L = 22$ and eight in the case with $L = 100$) fall outside the unit circle (Figs.~\ref{Fig:EV_KS}(b) and (d)), indicating that the dynamical system has some growing modes, the nonlinear part of the predictive method ($B \bs{f}$) suppresses the unbounded growth of these modes. This is fully compatible with the underlying physics of nonlinear dynamical systems, in which the unbounded growth of the unstable modes is suppressed by the energy-conserving nonlinear interactions that transfer energy to the stable modes where the dissipation occurs \citep{Sapsis2013, Majda2016, Qi2016}. The close match between the blue circles and red crosses in the right panels of Fig.~\ref{Fig:EV_KS} also shows that halving the length of training set does not change the eigenvalues identified by M2, confirming that these eigenvalues are captured robustly, and the growing modes are integral to the dynamical system. We also highlight that training in time-delay coordinate is a crucial part of M2 as choosing $q = 1$ for this method yields predictions that diverge from the actual data in less than $0.4/\Lambda_{max}$.

\subsection{Accurate representation of nonlinearity beyond squared terms \label{Section:NL}}
  
Here, we consider a different commonly-explored prototype for chaotic dynamics, Lorenz-96 system \citep{Lorenz2006}, whose governing ordinary differential equation (ODE) is given by 
\begin{eqnarray}
	\dot{X}_j = (X_{j+1} - X_{j-2})X_{j-1} - X_j + F  \, , \label{eqn:Lorenz96}  
\end{eqnarray}
where $X$ and overdot indicate the Lorenz variable and time derivative, respectively, and $j$ varies from $1$ to $n = 40$. The external forcing term $F$ determines the level of chaoticity. Here, we take $F = 16$. Again, RK4 along with periodic boundary condition is used to numerically integrate the system, and to build training sets with $N = 100000$ data points uniformly sampled at every $\tau = 0.02\tau_d$.

\begin{figure}
  \centerline{\includegraphics[width=1.0\textwidth]{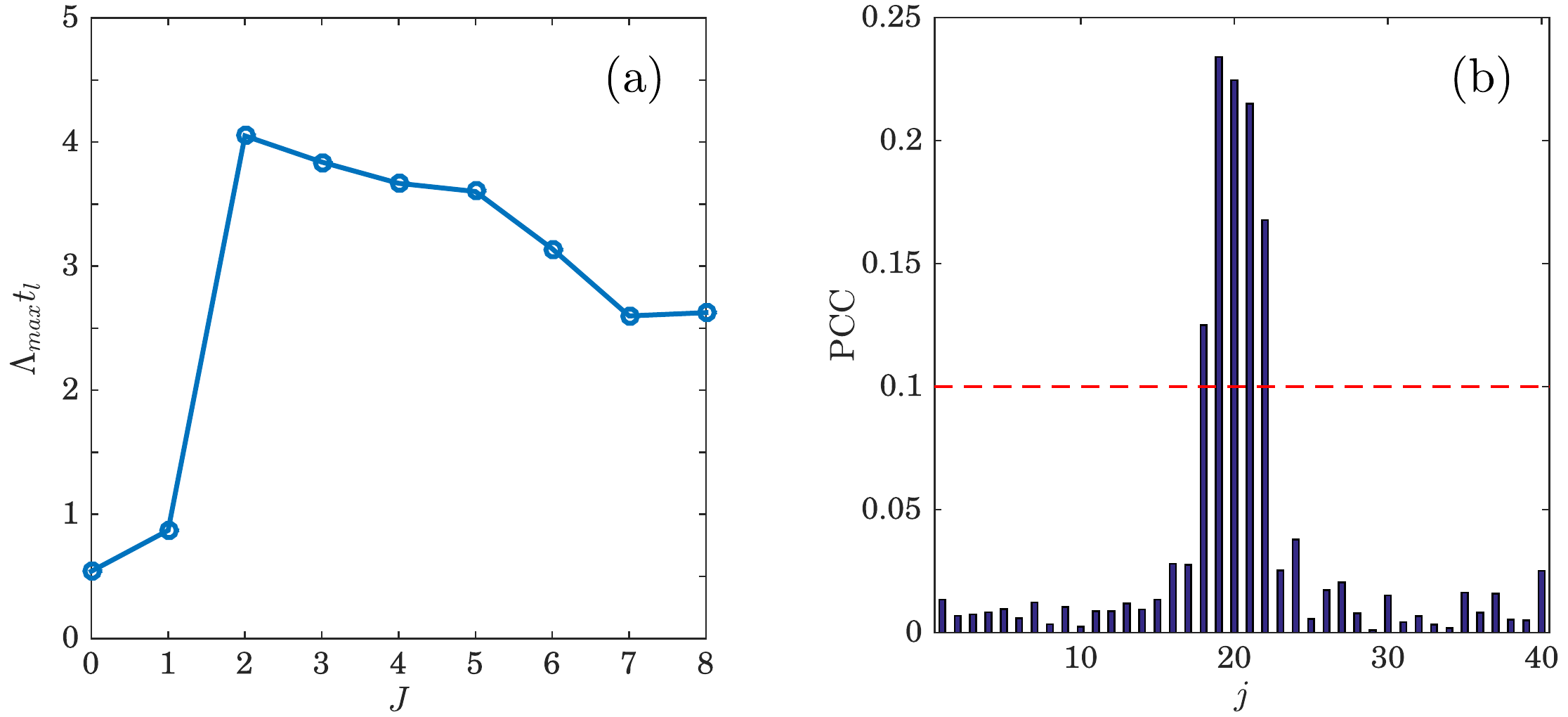}}
 \caption{(a) Variation of prediction horizon $t_l$ with the number of neighboring points $J$ used for constructing quadratic forcing terms. (b) Pearson correlation coefficient ($\mathrm{PCC}$) between the time series of the temporal derivative at a certain point ($\dot{X}_{20}$) and the time series of $X$ at every grid point. The horizontal dashed line corresponds to the threshold 0.1 for choosing the neighboring points. For the Lorenz-96 system under consideration, the number of grid points, external forcing term and attractor dimension respectively equal $n = 40$, $F = 16$ and $D_{KY} = 32$.}
\label{Fig:Lorenz_NL}
\end{figure}

For this test case, although the nonlinearity still has a quadratic form, constructing the forcing vector $\bs{f}$ using only $X_j^2$ yields short prediction horizons (see the point corresponding to $J = 0$ in Fig. \ref{Fig:Lorenz_NL}(a)). Seemingly, this is owing to the different nature of nonlinearities appearing in ODE (\ref{eqn:Lorenz96}) of Lorenz-96 compared to those in the K-S equation; where the former involves terms which are the product of $X$ at some distinct grid points, e.g. $X_{j-1} X_{j+1}$. Motivated by the underlying dynamics of Lorenz-96, we construct the forcing vector so that in addition to $X_j^2$, it involves terms in the form of $[X_jX_{j+1} \ X_jX_{j+2} \ \cdots \ X_jX_{j+J}]$, where $J$ indicates the number of `neighboring' points incorporated for building the forcing vector. The vector hence finds the following form     
\begin{eqnarray}
    \bs{f}^T = \left[
   \begin{array}{cccccccccccc}
    X_1^2      & 
    X_1X_2     & 
    \cdots     & 
    X_1X_{J+1} & 
   \cdots      & 
    X_j^2      & 
    X_jX_{j+1} & 
    \cdots     & 
    X_jX_{j+J} & 
    \cdots     & 
    X_nX_{n+J}  
   \end{array}  
\right]  \, .  \label{eqn:Forcing_Lorenz}
\end{eqnarray}
Note that when the neighboring points fall outside the domain, boundary conditions are invoked. 

Figure \ref{Fig:Lorenz_NL}(a) demonstrates how the prediction horizon $t_l$ changes as the number of neighboring points in the forcing vector grows. An abrupt jump in $t_l$ is observed when $J$ increases from 1 to 2, which is consisent with the underlying ODE of Lorenz-96 in which nonlinear terms in the form of $X_{j-1}X_{j+1}$ are present. Further increase in $J$ results in a gradual decline in $t_l$, as the size of forcing vector, and subsequently, its coefficient matrix $\mrm{B}$, unnecessarily grow, which leads to less accurate approximations of the components of this matrix.   

In the problems for which no knowledge of the underlying dynamics is available, a fully data-driven alternative approach for detecting the nonlinearities can be sought by calculating the Pearson correlation coefficients (PCC) between the time series of the temporal derivative at a certain point $\dot{X}_I$ and the time series of $X_{j}$ at all points as
\begin{eqnarray}
	\mrm{PCC} = \frac{\bs{E} \big[ (X_j - \mu_{X_j})(\dot{X}_I- \mu_{\dot{X}_I}) \big]}{s_{X_j} s_{\dot{X}_I}}  \, , \label{eqn:PCC}  
\end{eqnarray}
where $\mu$ and $s$ denote the mean and standard deviation of each time series, respectively, and $\bs{E}$ represents expecation operator. However, due to the chaoticity of the system, the temporal evolution of Lorenz variable at all grid points are interconnected, the dependence is anticipated to be stronger, when according to ODE (\ref{eqn:Lorenz96}), $\dot{X}_I$ is directly a fucntion of $X_{j}$. Consequently, as can be seen in Fig. \ref{Fig:Lorenz_NL}(b), PCC is substantially larger for grid point $I$ or the points in one- or two-grid-point distance from $I$, since terms involving these points explicitly appear in the underlying ODE of Lorenz-96. The points above the threshold (dashed red line in Fig. \ref{Fig:Lorenz_NL}(b)) can then be selected as the neighboring points while constructing the forcing vector. We remark that this data-driven approach solely identifies the neighboring points, and does not provide any information with regard to the order of nonlinearity. Once the neighboring points are detected, one can obtain low- to high-order monomials by multiplying the Lorenz variables at the neighboring points with each other, and stack up these terms to build the forcing vector. The highest-order term required for constructing the forcing vector can be determined by the investigator's speculation or intuition.

\color{black}

\section{Testing the performance of method M2 for various chaotic dynamical systems \label{Section:Results}}

In the following section, we present the detailed results of the Koopman-based method that treats the nonlinearities as external actuations (M2), when it is used to predict the spatiotemporal evolution of a variety of chaotic systems, from the simple and commonly-used Lorenz-63 to high-dimensional and highly chaotic K-S and Lorenz-96 systems. The section is then concluded by a more complex and larger-scale fluid example (a 2D lid-driven cavity flow at $\mrm{Re} = 20000$ and $30000$). For all examples, prediction horizon $t_l$ and averaged error $E_{ave}$ are calculated based on the definitions and procedures introduced in Sec.~\ref{Section:Theory}.  


\subsection{The Lorenz-63 system \label{Section:Lorenz63}} 

Lorenz-63, one of the most well-known prototypes of chaotic dynamics, was originally developed by \citet{Lorenz1963} as a simplified mathematical model for the atmospheric convection from the relatively complicated equations of motion and heat transport for an incompressible Boussinesq flow. The dynamics of the Lorenz-63 system is characterized by the following ODEs
\begin{eqnarray}
	\dot{x} &=& \sigma(y - x)  \, , \nonumber \\
	\dot{y} &=& x(\rho - z) - y  \, ,  \label{eqn:Lorenz63}  \\
	\dot{z} &=& xy - \beta z  \, ,  \nonumber
\end{eqnarray}
where $\sigma$ and $\rho$ represent the Prandtl and scaled Rayleigh numbers, respectively, and $\beta$ is related to the dimensions of the atmospheric layer. Following \citet{Lorenz1963}, we take $\sigma = 10$, $\rho = 28$ and $\beta = 8/3$, for which the system reveals a chaotic beahaviour with a strange, butterfly-like attractor. The same RK4 scheme as Sec.~\ref{Section:Theory} is used to numerically integrate ODEs (\ref{eqn:Lorenz63}), and to construct training sets with $100000$ data points, which are uniformly sampled at every $\tau = 0.02\tau_d$. Note that while developing a data-driven model via M2, all possible quadratic combinations of $x$, $y$ and $z$ are included in the forcing vector of Eq.~(\ref{eqn:M2}), i.e. $\bs{f}^T = [x^2 \ y^2 \ z^2 \ xy \ xz \ yz]$.        

\begin{figure}
  \centerline{\includegraphics[width=0.95\textwidth]{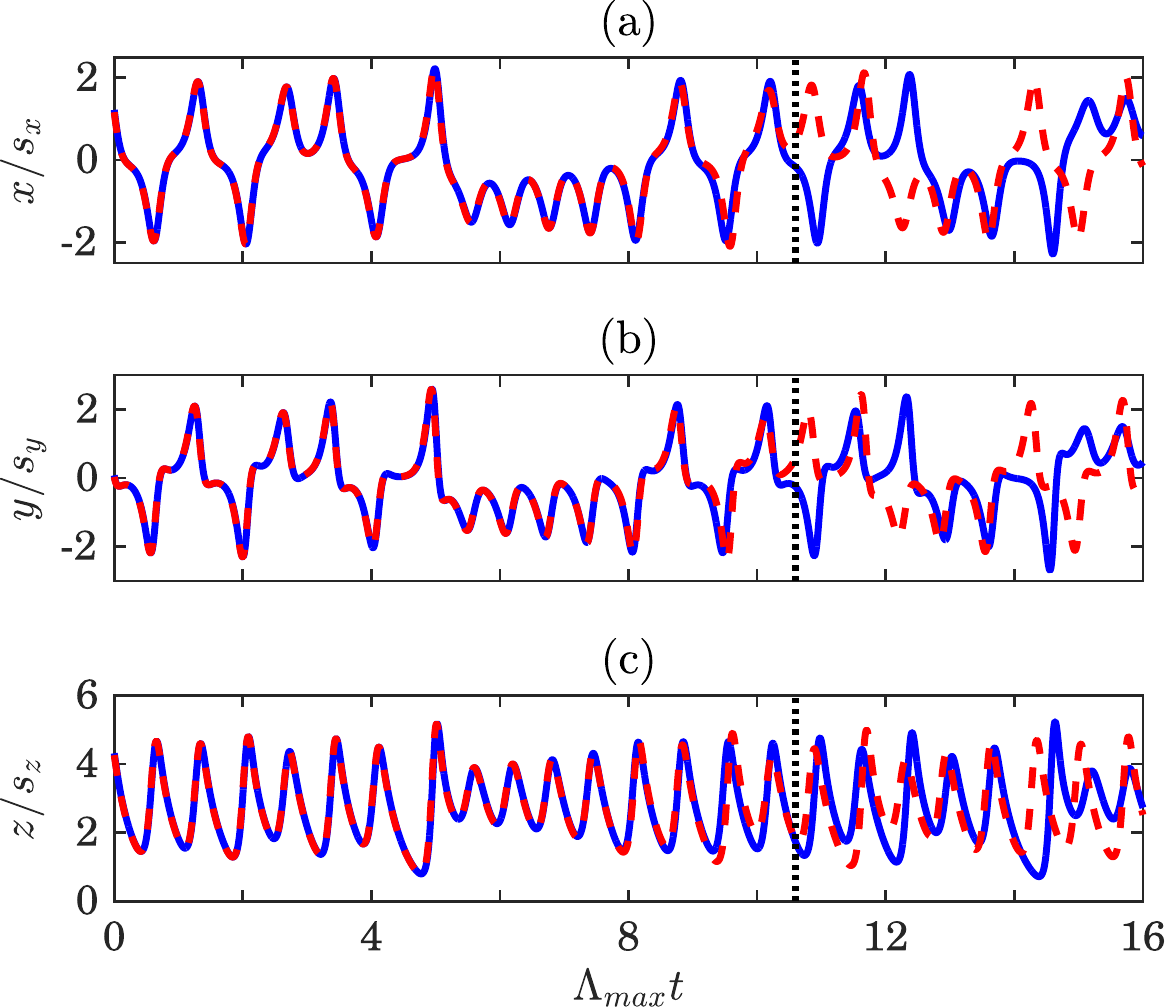}}
 \caption{Predictions of the present method M2 for the temporal evolution of the variables in a Lorenz-63 system with $\sigma = 10$, $\rho = 28$ and $\beta = 8/3$, where the method parameters are selected as $(r, p, q) = (60, 120, 50)$. For this Lorenz-63 system, the Kaplan-Yorke-based attractor dimension and the leading Lyapunov exponent of the system are found to be $D_{KY} = 2.06$ and $\Lambda_{max} = 0.91$. Note that vertical axes are normalized by the standard deviation of the corresponding variable from testing data, denoted by $s_x$, $s_y$ or $s_z$.}
\label{fig:Lorenz63}
\end{figure}

\begin{figure}
  \centerline{\includegraphics[width=1.0\textwidth]{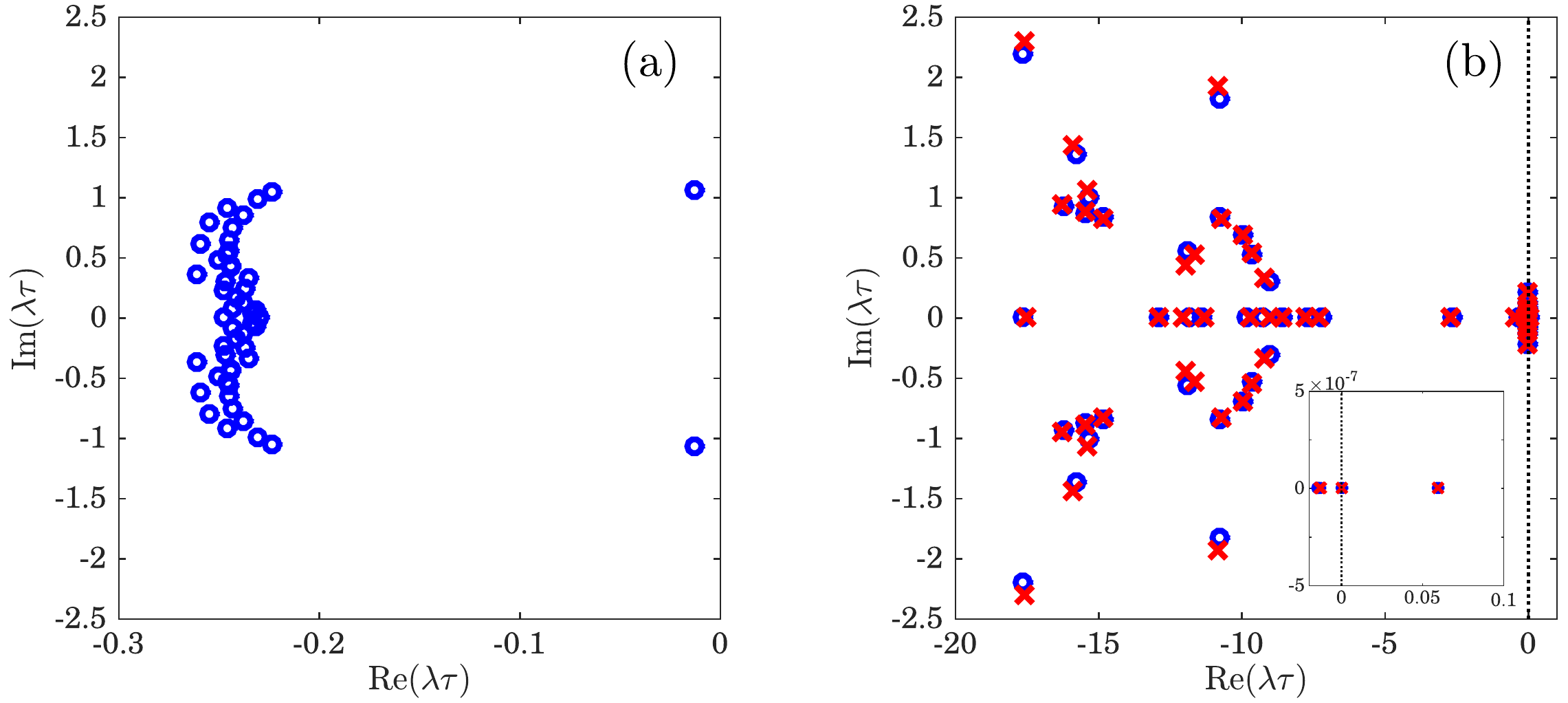}}
 \caption{Eigenvalues $\lambda$ of $\exp(\tau \mrm{A})$ for the under consideration Lorenz-63 system, scaled by sampling frequency $1/\tau$, and obtained using (a) M1 with $(r, q) = (100, 50)$, (b) M2 with $(r, p, q) = (60, 120, 50)$. The blue circles show the eigenvalues given by the entire training set with $100000$ samples, while the red crosses in panel (b) indicate the eigenvalues identified using half the training set. The close agreement between the two suggests that the eigenvalues of the system are captured robustly. The vertical dotted line in panel (b) corresponds to the imaginary axis. For clarity, only the first $50$ eigenvalues are depicted. Smaller panel within panel (b) magnifies the eigenvalue of M2 with positive real part. Note that except for the few leading eigenvalues, modes given by M2 decay at much faster rates than those given by M1.}
\label{Fig:EV_Lorenz63}
\end{figure}

As shown in Fig.~\ref{fig:Lorenz63}, the predictions of M2 follow the true trajectory for more than 10 Lyapunov timescales with $E_{ave} = 2.10\%$, while the extreme events, i.e. events at which $\lvert x \rvert > 2s_x$ and $\lvert y \rvert > 2s_y$, are also accurately captured. We reiterate that using M1 or M2 with $q = 1$ results in rapidly diverging predictions, so that for both $\Lambda_{max} t_l < 0.1$, suggesting that the delay-embedding of the measurements along with the incorporation of physics-driven frocings are vital for reasonably long-time accurate predictions. As expected, again, all the eigenvalues of the system identified by M1 fall inside the unit circle (Fig.~\ref{Fig:EV_Lorenz63}a), which is consistent with the quickly vanishing predictions of this method for Lorenz-63.


\subsection{Kuramoto-Sivashinsky equation \label{section:KS}}

\begin{figure}
  \centerline{\includegraphics[width=1.0\textwidth]{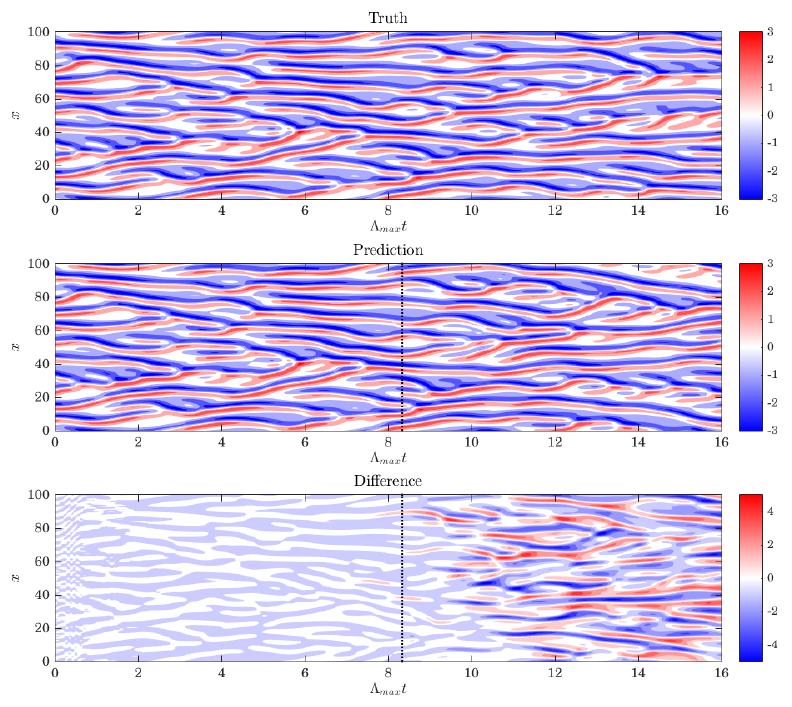}}
 \caption{Spatiotemporal evolution of an unforced K-S system with the domain length $L = 100$ and the Kaplan-Yorke dimension $D_{KY} = 23.2$. Shading shows (a) $u(x,t)$ from integration of Eq.~(\ref{eqn:KS}); (b) $u(x,t)$ predicted by M2 when a training set with $70000$ data points is used; (c) Difference of (a) and (b). The vertical dotted lines in panels (b) and (c) mark the divergence time $t_l$ of the data-driven predictions. $\Lambda_{max}$ is the largest positive Lyapunov exponent.}
\label{Fig:KS}
\end{figure}

The spatiotemporal prediction of M2 for a K-S equation with domain length $L = 100$ is displayed in Fig.~\ref{Fig:KS}(b), and is compared against the testing set provided by the numerical solver of Sec.~\ref{Section:Model} in Fig.~\ref{Fig:KS}(a). The difference of the two is shown in panel (c). Futhermore, the number of grid points $n$ needed for stably advancing the K-S equation (\ref{eqn:KS}) in time, the properties of the attractor ($D_{KY}$ and $\Lambda_{max}$), the method parameters $(r, p, q)$ leading to the best results, and the assessment of the method performance ($t_l$ and $E_{ave}$) are detailed in Table \ref{Table:KS}, for this case and several other K-S systems with different domain lengths. All cases in Table \ref{Table:KS} are identical to those examined in \citet{Pathak2018} in terms of domain length, potential forcing, attractor properties, and the length and sampling interval of the training set. As can be seen in this table, for moderately chaotic systems ($L \leq 200$), the data-driven predictions remain accurate for more than $8/\Lambda_{max}$, while $E_{ave}$ is below $7 \%$. As $L$ and choticity further grow, $t_l$ slowly declines and $E_{ave}$ gradually increases. Notwithstanding, for all considered cases, M2 provides skillfull forecasts for relatively long times, and it modestly outperforms the reservoir computing approach of \citet{Pathak2018} for which, $t_l$ was found to be around $6/\Lambda_{max}$ for all cases, when an adequeate number of parallel reservoirs had been used.

\begin{table}
\centering
\caption{Prediction horizon $t_l$ and averaged error $E_{ave}$ for K-S equation (\ref{eqn:KS}) at different levels of chaoticity controlled by domain length $L$. $n$ is the number of grid points or the physical dimension, and $D_{KY}$ denotes the attractor dimension calculated based on the Kaplan-Yorke formulation. Method parameters $(r, p, q)$ giving the best results are also reported. All cases are unforced, except those with asterisks for which $\eta = 0.01$.} 
\begin{ruledtabular}
\centering
\begin{tabularx}{\textwidth}{cccccc}
  \hline
   $L$    &  $n$ & $D_{KY}$ &      $(r, p, q)$     & $\Lambda_{max} t_l$ & $ E_{ave} $ \\
  \hline
  22      &  64  &   5.20   &   $(400, 400, 20)$   &         8.35        &    3.36     \\ 
  100     &  128 &   23.2   &   $(800, 800, 10)$   &         8.30        &    5.73     \\ 
  100$^*$ &  128 &   24.1   &   $(800, 800, 10)$   &         8.03        &    3.81     \\ 
  200     &  256 &   43.3   &  $(1200, 1200, 10)$  &         8.17        &    4.82     \\ 
  200$^*$ &  256 &   44.8   &  $(1200, 1200, 10)$  &         8.17        &    6.95     \\ 
  400     &  256 &   85.0   &  $(3000, 3000, 15)$  &         7.51        &    6.12     \\ 
  800     &  512 &   167    &  $(6000, 6000, 15)$  &         7.12        &    7.32     \\ 
  1600    & 1024 &   338    & $(16000, 16000, 20)$ &         6.43        &    10.9     \\ \hline
\end{tabularx}
\end{ruledtabular}
\label{Table:KS}
\end{table}


\subsection{The Lorenz-96 system \label{Section:Lorenz96}} 

\begin{figure}
  \centerline{\includegraphics[width=1.0\textwidth]{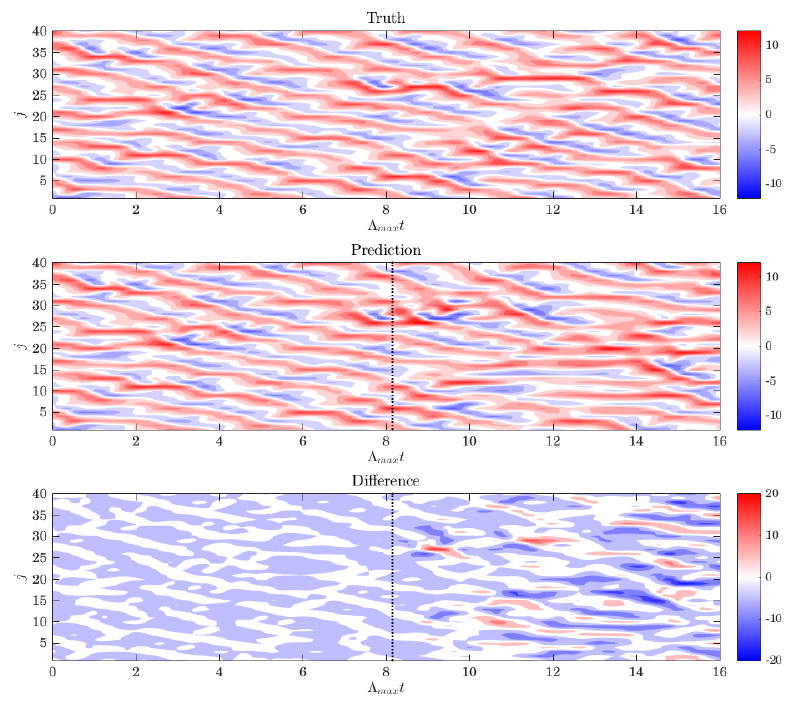}}
 \caption{Spatiotemporal evolution of a Lorenz-96 system with the external forcing $F = 8$ and the Kaplan-Yorke dimension $D_{KY} = 28.4$. (a) Testing set obtained by numerically intergrating Eq.~(\ref{eqn:Lorenz96}); (b) M2 predictions while the model is built using a training set with $100000$ samples; (c) Difference of (a) and (b). The vertical dotted lines in panels (b) and (c) correspond to the time $t_l$ at which the predictions of the data-driven method diverge from the actual data. $\Lambda_{max}$ is the largest positive Lyapunov exponent.}
\label{Fig:Lorenz96}
\end{figure}

Figure \ref{Fig:Lorenz96} depicts the spatiotemporal evolution of a Lorenz-96 system with the external forcing $F = 8$, given by RK4 integration of Eq.~(\ref{eqn:Lorenz96}) (panel (a)) and the data-driven prediction of M2 (panel (b)). The last panel shows the difference between the two results. In addition to this case, two other systems with $F = 4$ (lower chaoticity) and $F = 16$ (higher chaoticity) are explored as well. For all cases, the number of grid points is fixed at $n = 40$. Further details of each case, viz., attractor properties ($D_{KY}$ and $\Lambda_{max}$), the trio of optimal method parameters $(r, p, q)$, the prediction horizon $t_l$ and the averaged error $E_{ave}$ are reported in Table~\ref{Table:Lorenz96}. It is observed that when the system exhibits a quasiperiodic behavior and has a power spectrum with some local maxima (e.g. the test case with $F = 4$), contingent upon the availability of enough snapshots for training, M2 predictions can be accurate for very long times, and occassionally they may never diverge, meaning the underlying dynamics can be fully discovered. As the chaoticity of the Lorenz-96 system is increased by doubling $F$, $t_l$ becomes finite ($8.16/\Lambda_{max}$) and $E_{ave}$ grows by almost $2.5 \%$. Further doubling of $F$ makes the system highly chaotic so that forecasting its spatiotemporal evolution becomes challenging. Nevertheless, the present Koopman-based method still yields predictions which are accurate for more than $4/\Lambda_{max}$ with $E_{ave} = 7.67 \%$, when $F = 8$.  

\begin{figure}
  \centerline{\includegraphics[width=1.0\textwidth]{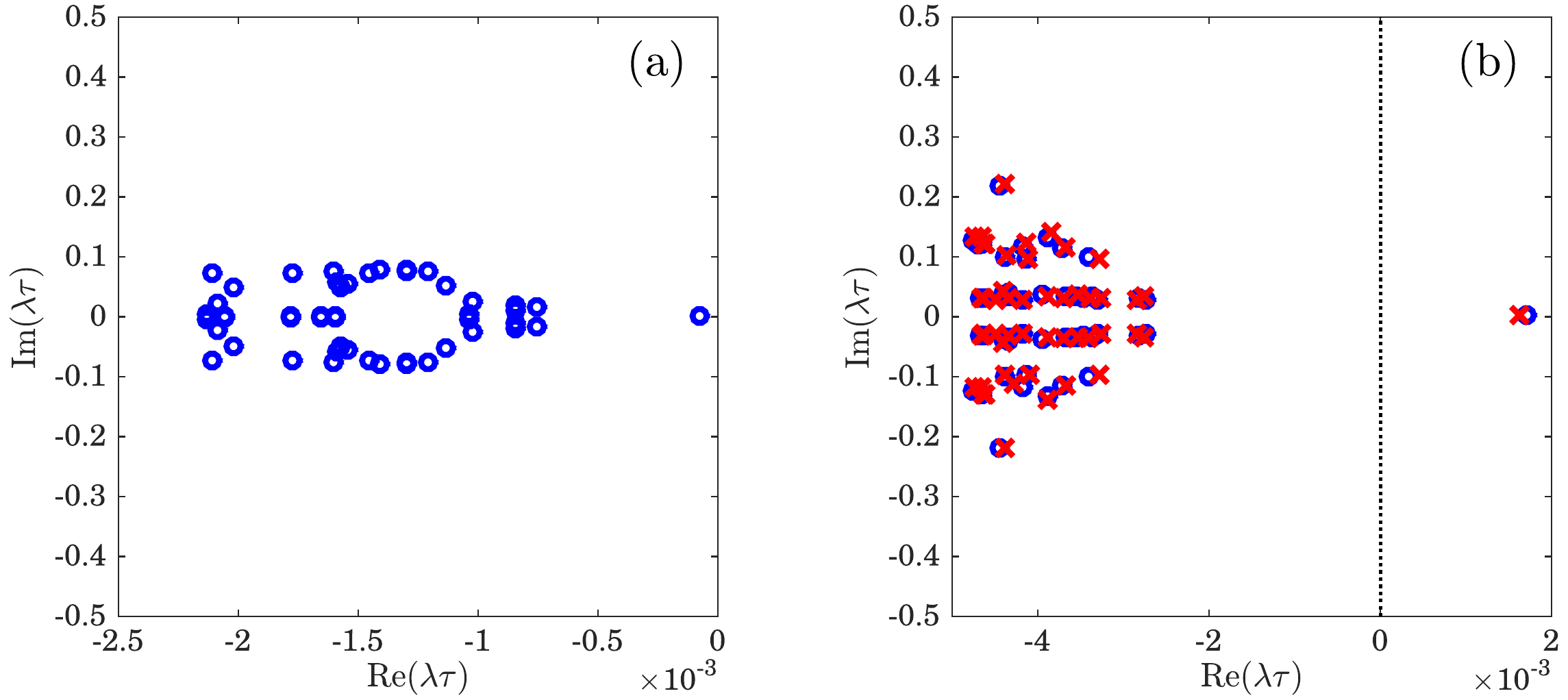}}
 \caption{Eigenvalues $\lambda$ of $\exp(\tau \mrm{A})$ for the studied Lorenz-96 system with $F = 8$, scaled by sampling frequency $1/\tau$, and provided by (a) M1 with $(r, q) = (200, 20)$, (b) M2 with $(r, p, q) = (700, 2100, 20)$. The blue circles and red crosses show the eigenvalues detected by the entire and half the training set, respectively. The close agreement between the two suggests that the eigenvalues of the system are captured robustly. Vertical dotted lines in panel (b) indicates the imaginary axis. For clarity, only the first $50$ eigenvalues are depicted. Smaller panel within panel (b) magnifies the eigenvalues of M2 corresponding to the growing modes. Note that except for the few leading eigenvalues, modes given by M2 decay at much faster rates than those given by M1.}
\label{Fig:EV_Lorenz96}
\end{figure}

Figure \ref{Fig:EV_Lorenz96} displays the eigenvalues of the Lorenz-96 system with $F = 8$ given by M1 (left) and M2 (right). As demonstrated by this figure, not surprisingly, all modes of $\mrm{A}_{\mrm{HDMD}}$ are decaying again, leading to the predictions that approach zero fairly rapidly, and lose the true trajectory in less than $0.1/\Lambda_{max}$. Furthermore, delay-embedding of the vector-valued observables was found to be essential in M2, so that choosing $q = 1$ using this method resulted in $t_l < 0.2/\Lambda_{max}$.

\begin{table}
\centering
\caption{Prediction horizon $t_l$ and averaged error $E_{ave}$ for Lorenz-96 system (\ref{eqn:Lorenz96}) with different external forcings $F$. $n$ and $D_{KY}$ indicate the number of grid points and Kaplan-Yorke-based attractor dimension, respectively. Method parameters $(r, p, q)$ leading to the most accurate predictions are also presented. Despite using very long testing sets ($\sim 30/\Lambda_{max}$), predictions for the case with $F = 4$ were seen to agree very closely with the actual data for the entire length of testing sets so that the prediction error always remained below the divergence threshold.} 
\begin{ruledtabular}
\centering
\begin{tabularx}{\textwidth}{cccccc}
  \hline
   $F$ &  $n$ & $D_{KY}$ &      $(r, p, q)$    & $\Lambda_{max} t_l$   & $ E_{ave} $ \\
   \hline
    4  &  40  &   15.3   &  $(400, 1200, 20)$  & No divergence observed &    4.31     \\ 
    8  &  40  &   28.4   &  $(700, 2100, 20)$  &            8.16        &    6.82     \\ 
    16 &  40  &   32.1   &  $(1200, 1200, 40)$ &            4.05        &    7.67     \\ \hline
\end{tabularx}
\end{ruledtabular}
\label{Table:Lorenz96}
\end{table}


\subsection{2D lid-driven cavity flow \label{Section:Cavity}} 

\begin{figure}
  \centerline{\includegraphics[width=0.45\textwidth]{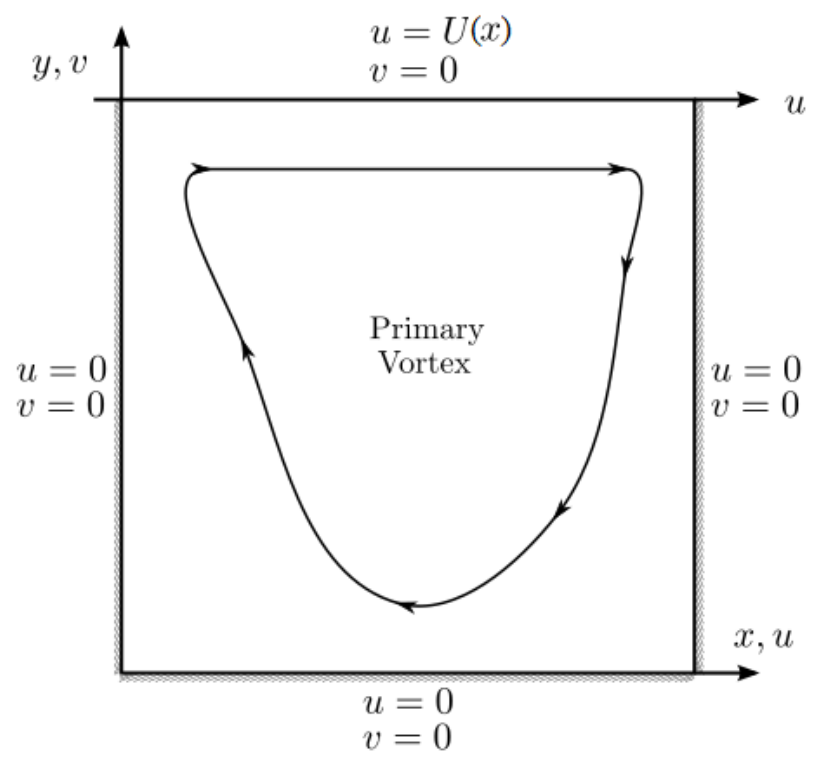}}
 \caption{Schematic of a 2D lid-driven cavity flow. Here, $u$ and $v$ are the horizontal and vertical velocities, respectively, and $U(x) = 16x^2(1-x)^2$.}
\label{Fig:Cavity}
\end{figure}

The 2D lid-driven cavity flow has been employed for decades as a benchmark for validation of new numerical models and computational schemes (See e.g. \cite{Ghia1982, Schreiber1983, Sahin2003}). Here, we choose this problem as a gateway to the implementation of our Koopman-based method to large-scale fluid flows at high Reynolds numbers. The schematic of the 2D cavity flow is sketched in Fig.~\ref{Fig:Cavity}. The constant-density fluid is confined by a square box whose walls are stationary, except for the top wall (lid), which moves to the right with the velocity $U(x)$. This produces a shear-driven flow mixing the entire fluid via the clockwise primary vortex at the center, as well as some smaller-scale vortices at the corners, if Reynolds number is sufficiently large. The nondimensional equations of motion for this unsteady and incompressible flow are in the following form
\begin{eqnarray}
	\frac{\partial u}{\partial x} + \frac{\partial v}{\partial y} &=& 0 \, , \nonumber \\
	\frac{\partial u}{\partial t} + u\frac{\partial u}{\partial x} + v\frac{\partial u}{\partial y} &=& - \frac{\partial p}{\partial x} + \frac{1}{\mrm{Re}} \bigg(\frac{\partial^2 u}{\partial x^2} + \frac{\partial^2 u}{\partial y^2} \bigg)  \, , \\
	\frac{\partial v}{\partial t} + u\frac{\partial v}{\partial x} + v\frac{\partial v}{\partial y} &=& - \frac{\partial p}{\partial y} + \frac{1}{\mrm{Re}} \bigg(\frac{\partial^2 v}{\partial x^2} + \frac{\partial^2 v}{\partial y^2} \bigg)  \, , \nonumber
\label{eqn:Cavity}  
\end{eqnarray}
where $u$, $v$ and $p$ denote the horizontal velocity, the vertical velocity and the pressure fields, respectively. Characteristic length and velocity are selected as the domain length $L$ and maximum lid velocity $U_{max}$, so that $\mrm{Re} = U_{max} L/ \nu$, with $\nu$ indicating the kinematic viscosity of the fluid. This also means that the time is nondimensionalized by the advective timescale $L/U_{max}$. No-slip boundary conditions are enforced along all walls, except for the lid at which
\begin{eqnarray}
	U(x) = 16x^2(1-x)^2 \, , \, v = 0 \, . \label{eqn:BC}
\end{eqnarray} 
This boundary condition allows for smooth transitions in top corners, while satisfying the continuity and incompressibility. 

\begin{figure}
  \centerline{\includegraphics[width=0.5\textwidth]{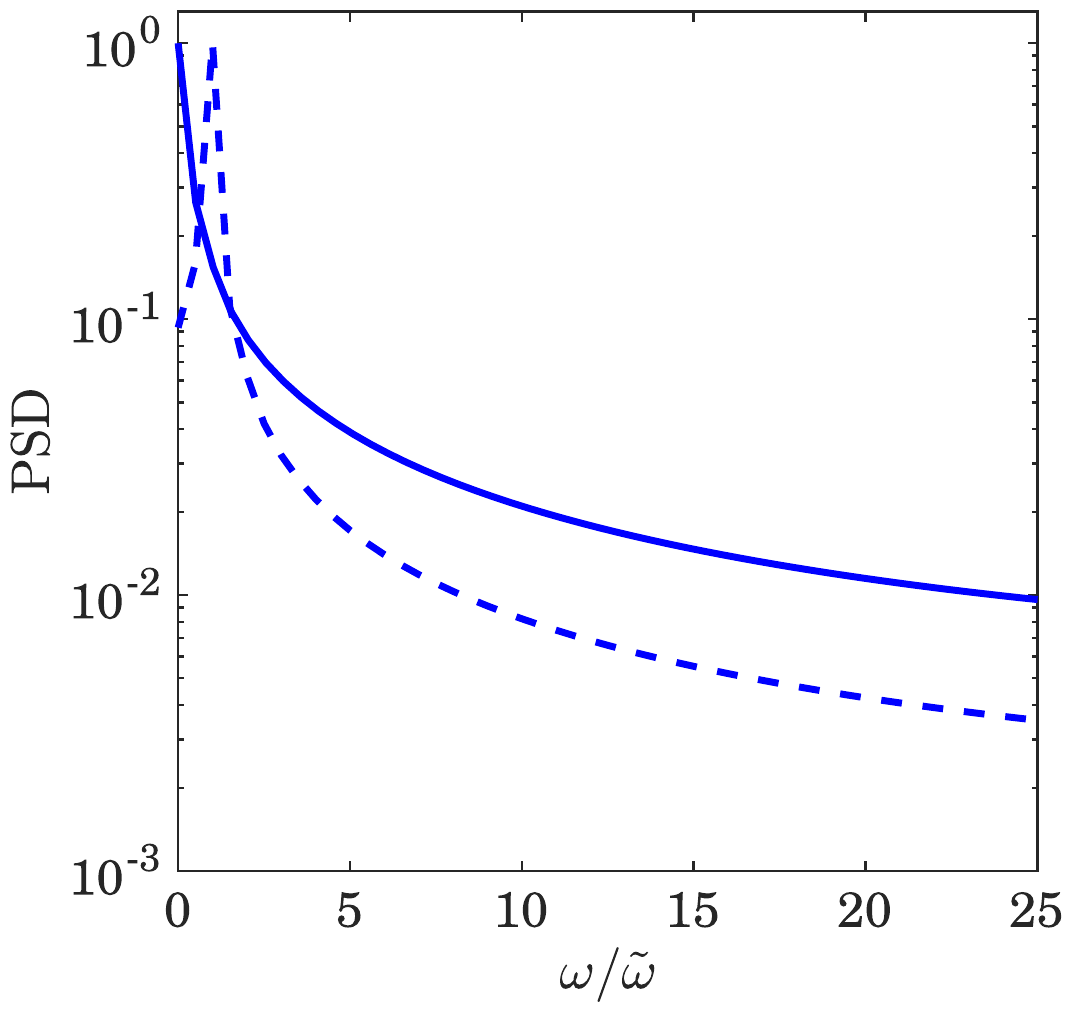}}
 \caption{The Power Spectral Densities (PSDs) of the time series of the PC1 calculated for the velocity field at $\mrm{Re} = 20000$ (dashed) and $\mrm{Re} = 30000$ (solid). The PSDs are evaluated by dividing the entire data consisting of $100000$ data points ($\sim 2000\tau_{adv}$) into 200 windows with the same length, and carrying out fast Fourier transform (FFT) coupled with Hann filter for each window. The results of all windows are then averaged to obtain the plotted PSDs. Frequency $\omega$ is normalized by the frequency of the advective timescale $\tilde{\omega} = 2\pi/\tau_{adv}$.}
\label{Fig:Spectra}
\end{figure}

To construct the required training and testing sets, direct numerical simulations (DNS) of the flow based on primitive variables are conducted using a Chebyshev-Chebyshev pseudospectral solver with $70$ grid points in each direction, and dimensionless time-step $\Delta t = 0.005$. The sufficiency of the number of grid points was examined via mesh refinement. The simulations are then carried out to build the model for the fluid system at two relatively high Reynolds numbers $\mrm{Re} = 20000$ and $\mrm{Re} = 30000$. We then sample horizontal and vertical velocities at every other grid point (in total $35$ grid points in each direction). The sampling interval and the length of training set for both cases are taken as $\tau \approx 0.025\tau_d$ and $T_{train} = 2000\tau_{adv}$. We then remove their corresponding long time-mean values, $\bar{u}$ and $\bar{v}$, to obtain anomalous velocities $u' = u - \bar{u}$ and $v' = v - \bar{v}$, and finally arrange these mean-removed velocities in 1D vectors $\bs{u'}$ and $\bs{v'}$. The state vector $\bs{V}$ at each time consists of these two vectors stacked on top of each other, i.e. $\bs{V} = [\bs{u} \ \bs{v}]^T$. Motivated by the underlying physics of the flow, we choose the Reynolds stress terms at each sampled grid point (${u'}^2$, $u'v'$ and ${v'}^2$) to form the nonlinear forcing vector $\bs{f}$.  The results are reported for the normal-to-the-plane vorticity field $\omega = \partial v /\partial x - \partial u /\partial y$, and the PC1 timeseries of the horizontal velocity. 

As discussed in \citet{Arbabi2017}, for $\mrm{Re} \leq 10000$, the cavity flow converges to a steady-state laminar solution whose corresponding attractor is in the form of a fixed point. Upon slight increase in $\mrm{Re}$ above 10000, a periodic flow with a single oscillation frequency emerges. This behavior persists until at $\mrm{Re} \geq 15000$ another bifurcation occurs and a flow with quasiperiodic behavior (multiple basic frequencies) forms. The third bifurcation occurs around $\mrm{Re} = 18000$ leading to a rapid rise in the level of kinetic energy. The kinetic energy then continually increases so that at $\mrm{Re} \gtrsim 22000$, the fluid system becomes fully chaotic and no quasiperiodic compoenents can be further detected. Figure \ref{Fig:Spectra} is consistent with the findings of \citet{Arbabi2017} so that for the cavity flow at $\mrm{Re} = 20000$, the power spectrum calculated for the timeseries of PC1 of the velocity field has a maximum at $\omega/\tilde{\omega} \approx 1$, manifesting that the underlying dynamics is not fully chaotic yet. In fact, according to \citep{Arbabi2017}, such flow has a mixed spectrum, i.e. it contains both discrete and continuous components. In contrast, flow at $\mrm{Re} = 30000$ exhibits a monotonically decaying spectrum, which is indicative of chaotic behaviour.    

\begin{figure}
  \centerline{\includegraphics[width=1.0\textwidth]{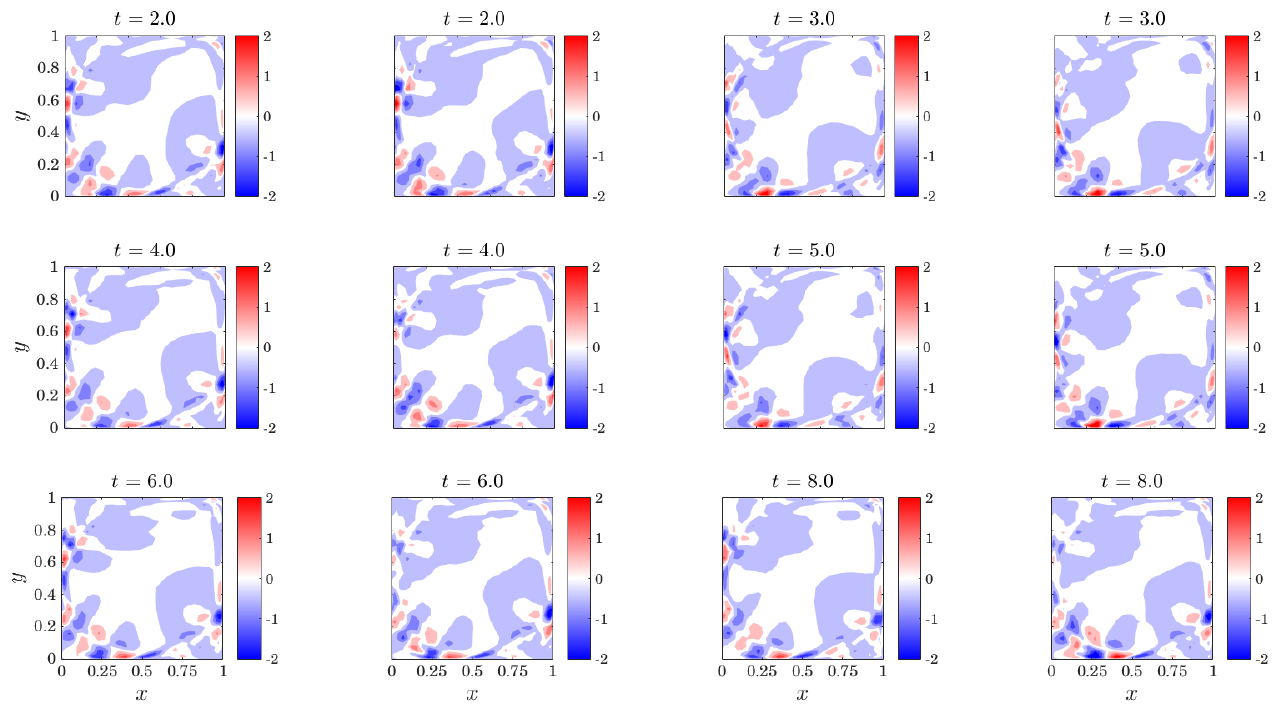}}
 \caption{Snapshots of vorticity field for a lid-driven cavity flow at $\mrm{Re} = 20000$ at six distinct dimensionless times given by DNS solver (first and third columns) and the best predictions of M2 (second and fourth columns). These results were obtained by choosing $(r, p ,q) = (400, 600, 3)$, leading to $t_l = 5.2\tau_{adv} = 1040 \Delta t$ and $E_{ave} = 16.6 \%$. Note that for the bottom panels, data-driven predictions have already deviated from DNS data.}
\label{Fig:Re20K}
\end{figure}

\begin{figure}
  \centerline{\includegraphics[width=0.95\textwidth]{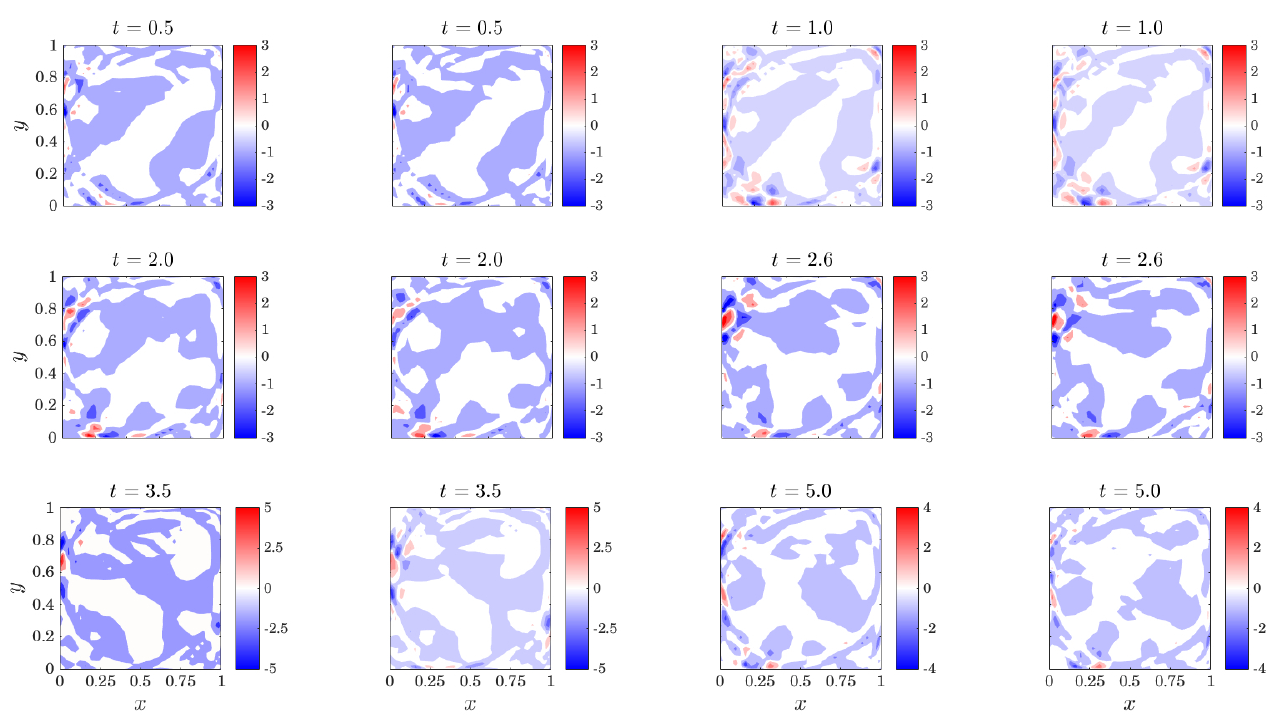}}
 \caption{Similar to Fig. \ref{Fig:Re30K}, but for a flow at $\mrm{Re} = 30000$. The optimal method parameters are found to be $(r, p, q) = (1200, 1800, 3)$, for whcih $t_l = 2.7\tau_{adv} = 540 \Delta t$ and $E_{ave} = 16.8 \%$.}
\label{Fig:Re30K}
\end{figure}

\begin{figure}
  \centerline{\includegraphics[width=0.95\textwidth]{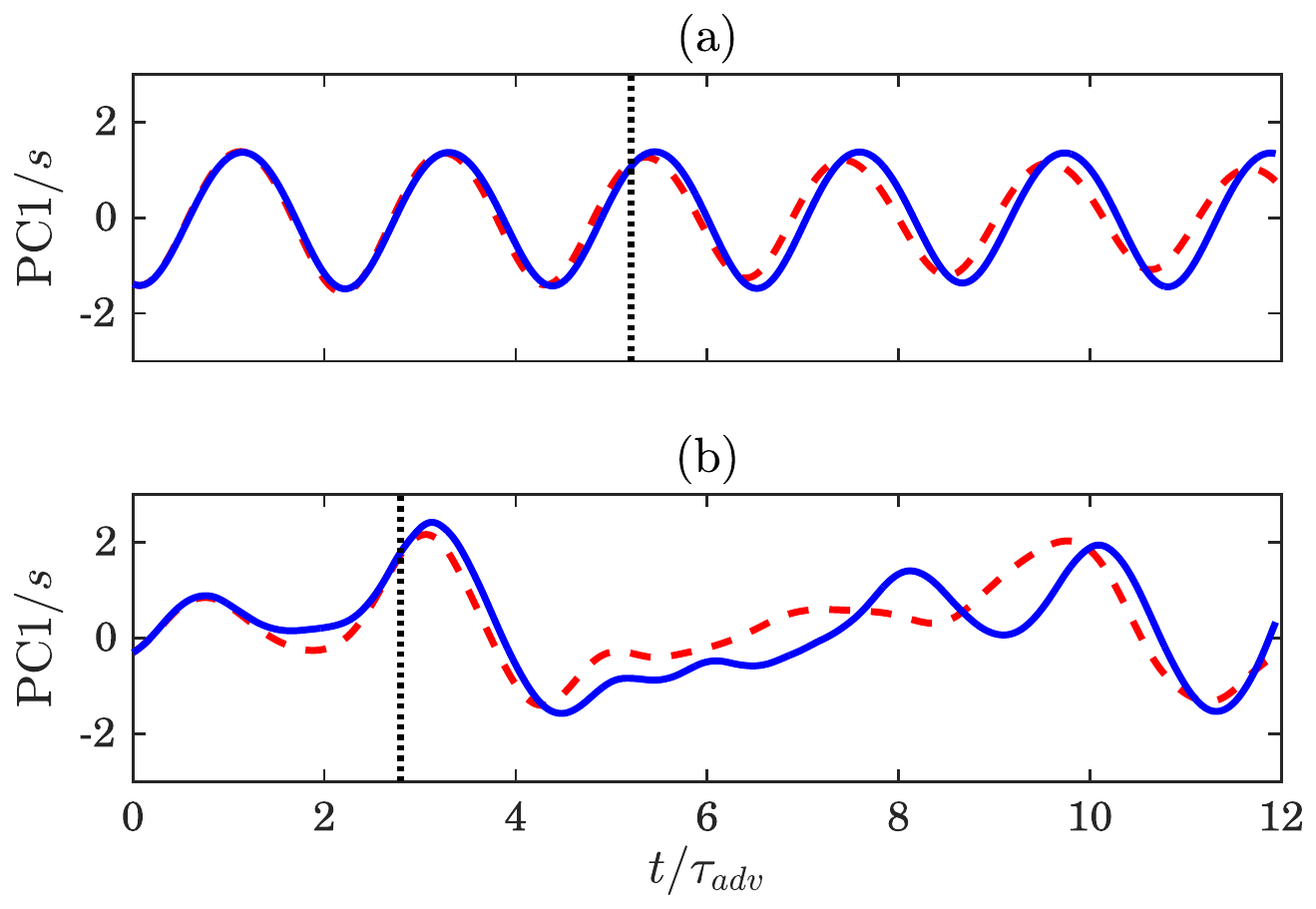}}
 \caption{Timeseries of the PC1 of horizontal velocity, calculated from DNS (solid blue) and predicted by M2 (dashed red), at $\mrm{Re} = 20000$ (top) and $\mrm{Re} = 30000$ (bottom). The results are scaled by the standard deviation $s$ of the PC1 timeseries from testing data. The vertical dotted lines correspond to the time at which the predictions of M2 for the entire flowfield (not merely the PC1 of $u$) diverge from DNS data.}
\label{Fig:PC1}
\end{figure}

As demonstrated by Figs.~\ref{Fig:Re20K} and \ref{Fig:Re30K}, the Koopman-based method M2 renders accurate predictions for several advective timescales, equivalent to hundreds of DNS time-steps (see the captions of the figures for the exact values), during which the vorticity fields unergo substantial changes. The presence of quasiperiodic components in the flow at $\mrm{Re} = 20000$ is distinctly illustrated by Fig. \ref{Fig:Re20K}, e.g. compare the snapshots of the vorticty field at $t = 2$ and $t = 4$ (or at $t = 3$ and $t = 5$), or see the PC1 timeseries shown in Fig.~\ref{Fig:PC1}(a). The performance of M2 somewhat degrades from $\mrm{Re} = 20000$ to $\mrm{Re} = 30000$ as the underlying dynamics becomes more complex, so that $t_l$ is nearly halved, while $E_{ave}$ varies insignificantly. Exact values are provided in the captions of corresponding figures. These observations are further supported by Fig. \ref{Fig:PC1} in which the timeseries for the PC1 of horizontal velocity obtained from DNS data (solid blue) and M2 predictions (dashed red) are displayed. Nonetheless, the results of M2 for the studied fluid example seem promising, and can lead to a new avenue for predictive modeling of high-dimensional and highly turbulent flows, specifically if some modifications are considered. This prospect will be discussed in detail in Sec.~\ref{Section:Conclusion}. 

\begin{figure}
  \centerline{\includegraphics[width=0.95\textwidth]{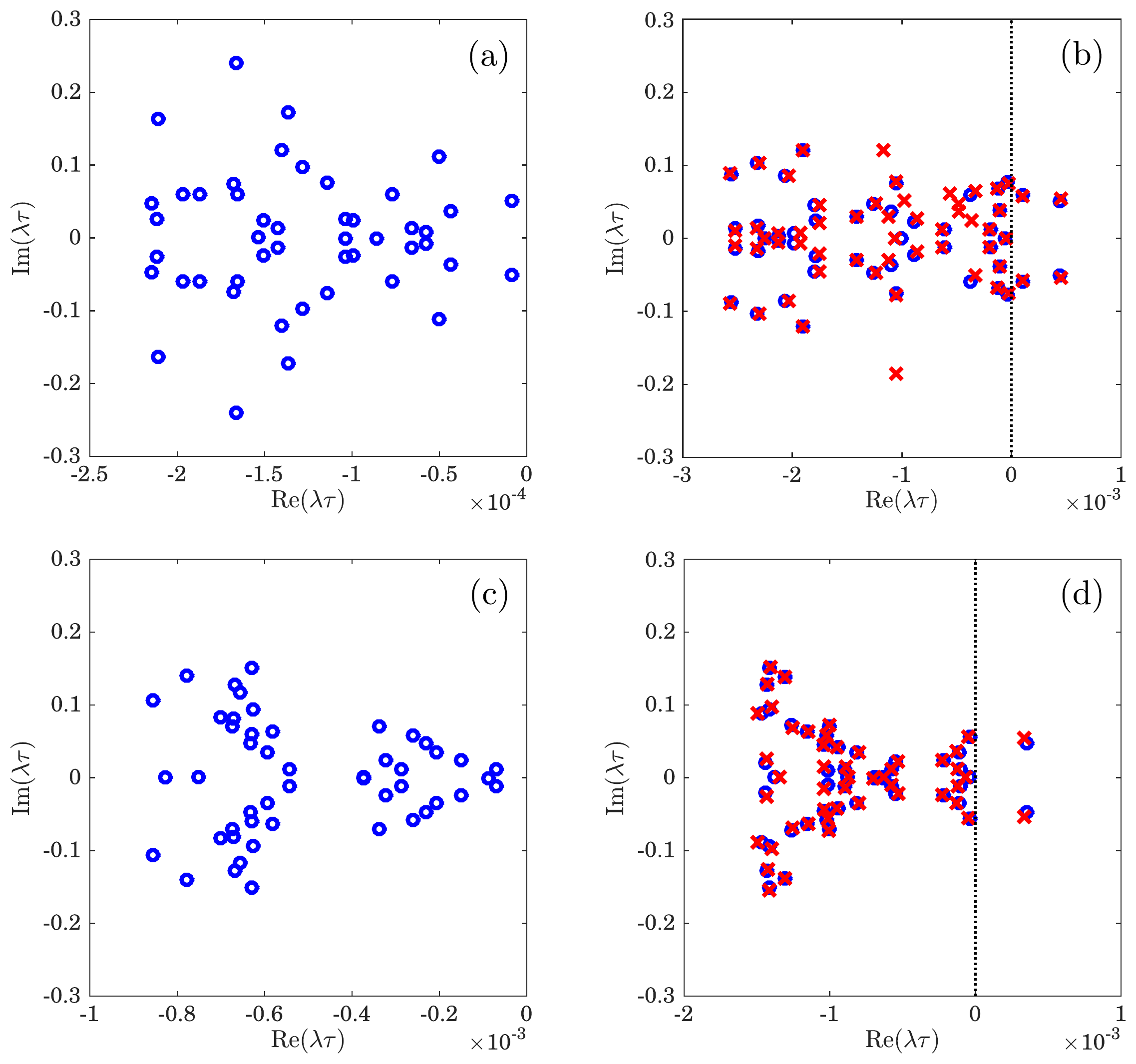}}
 \caption{Eigenvalues $\lambda$ of $\exp(\tau \mrm{A})$ for the described 2D cavity flow at $\mrm{Re} = 20000$ (top) and $\mrm{Re} = 30000$ (bottom), normalized by the sampling frequency $1/\tau$, and calculated via (a) M1 with $(r, q) = (1000, 10)$, (b) M2 with $(r, p, q) = (400, 600, 3)$, (c) M1 with $(r, q) = (1000, 10)$, and (d) M2 with $(r, p, q) = (1200, 1800, 3)$. The blue circles show the eigenvalues provided by the entire training set with the length $T_{train} = 2000\tau_{adv}$, whereas the red crosses in panels (b) and (d) exhibit the eigenvalues obtained using half the training set. The close agreement between the two suggests that the eigenvalues of the systems are captured robustly. Vertical dotted lines in these panels mark the imaginary axis. For clarity, only the first $50$ eigenvalues are displayed.}
\label{Fig:EV_Cavity}
\end{figure}

Similar to chaotic dynamical systems such as K-S equation and Lorenz-96, M2 discovers a few growing modes for the 2D cavity flow, irrespective of its Reynolds number (Figs. \ref{Fig:EV_Cavity}(b) and (d)), whose growths are suppressed by the existing nonlinearites as discussed in Sec.~\ref{Section:Model}. Moreover, M1 yields rapidly vanishing predictions which diverge from the actual flowfield in less than $40\Delta t$ when $\mrm{Re} = 20000$ and $20\Delta t$ when $\mrm{Re} = 30000$, as all eigenvalues detected by this method lie to the left of imaginary axis (Figs. \ref{Fig:EV_Cavity}(a) and (b)). Note also that the delay-embedding dimension chosen for this problem may seem small, but in fact $q\tau \approx 0.2\tau_d$, which is in the same range as what was found for the dynamical systems discussed earlier in this section. 

\color{black}

\section{Conclusions \label{Section:Conclusion}}

Within the present investigation, we have proposed a data-driven Koopman-based method, referred to as M2 in the text, which treats the nonlinearities of the system as external actuations, whereas the observables are linear, vector-valued, and time-delay-embedded. Hence, a linear framework (\ref{eqn:M2}) is built whose unknown maps are found via the DMDc technique of \citet{Brunton2016a}. This data-driven predictive framework is shown to accurately forecast the spatiotemporal evolution of common examples of chaos such as Lorenz-63, K-S and Lorenz-96 systems as well as a high-Reynolds-number fluid flow, namely, a 2D lid-driven cavity flow, for several Lyapunov or advective timescales, which, in the case of cavity flow, is equivalent to hundreds of numerical solver time-steps. 

As shown in Secs.~\ref{Section:Theory} and \ref{Section:Results}, the strong performance of M2 hinges on the simultaneous use of vector-valued, delay-embedded observables and physics-inspired forcings. The resulting linear model is built using the DMDc algorithm of \citet{Brunton2016a}. The advantages of delay-embedding had been shown in previous studies \citep{Giannakis2012, Tu2014, Brunton2017, Giannakis2017, Arbabi2017, Korda2018, Arbabi2018}. The novelty of M2 is in the last feature, i.e., the accurate representation of the underlying nonlinear processes using a linear model that treats the nonlinearities as exogenous forcings, which builds on the work of \citet{Brunton2017}. Note that in M2, the forcing terms are updated as the predictions of the new (future) state become available. Such representation of nonlinearities enables us to capture the potentially present unstable modes, whose unbounded growth is suppressed by the energy-conserving nonlinear interplay between the unstable and stable modes \citep{Sapsis2013, Majda2016, Qi2016}. It is worth noting that attempts on including such unstable modes in a linear model such as M1 (\ref{eqn:Pred}) leads to predictions that grow exponentially unboundedly, while attempts on excluding them, which are integral to the spatiotemporal evolution of the system, leads to inaccurate predictions. M2 provides a linear framework for accurately accounting for these unstable modes. 

Similar to most data-driven methods, and specifically for very large-scale and high-dimensional systems, the success of the present method to some degree depends on the availability of sufficiently long training sets. This is at least computationally very demanding, if not prohibitive, for three-dimensional and highly turbulent flows. Motivated by the success of \citet{Mohan2019} in accurately reproducing the long-term statistics of isotropic turbulence, we speculate that this issue might be rectified by initially compressing the 3D turbulence data via methods such as autoencoders from machine learning. The compressed data can then be used for training by M2, whose predictions can later be decoded and brought back to the physical space. \citet{Otto2019} have also shown that a neural network combining an autoencoder with linear recurrent dynamics can be employed to provide a low-dimensional dictionary of linear and nonlinear observales for the approximation of Koopman operator. The method of \citep{Otto2019} has proven to be skillful in identifying the salient dynamical modes, and making short-term predictions for some well-known chaotic dynamics. Successful implementation of these ideas can result in data-assisted surrogate models for computational fluid dynamics (CFD) solvers, that significantly reduce the computational time of these solvers by accelerating the advancement of the flow in time. We aim to pursue these lines of research in the subsequent sudies.       
 
\color{black}

\section*{Acknowledgment \label{section:Acknowledgment}}

We thank Hassan Arbabi for helpful discussions, insightful comments, and sharing his 2D incompressible flow DNS solver presented in \citep{Arbabi2017study}, Matthias Heinkenschloss, Igor Mezi{\'c}, and Ashesh Chattopadhyay for fruitful discussions, and Ashesh Chattopadhyay and David Lee for providing useful comments on the manuscript. We gratefully acknowledge the financial support from the NASA grant 80NSSC17K0266, a Faculty Initiative Fund award from the Rice University Creative Ventures, and the Mitsubishi Electric Research Labs (to P.H.), and NSF grant CCF-1816219 (to A.C.A.). This work used the Extreme Science and Engineering Discovery Environment (XSEDE) Stampede2 through allocation ATM170020, the Yellowstone high-performance computing system provided by NCAR's Computational and Information Systems Laboratory through allocation NCAR0462, and the DAVinCI cluster of the Rice University Center for Research Computing. Examples of codes and data can be found at \textcolor{blue}{\url{https://github.com/mkhodkar70/Short_term_forecast}}.

\color{black}

\appendix
\section{Methodical selection of Koopman-based method's parameters \label{Section:Parameter}}

For each case, the optimal choices of delay-embedding dimension $q$ and the size of the reduced subspace $r$ are obtained by a comprehensive search over a broad range of these parameters on validation sets, i.e. datasets which are fully independent from training or testing sets, and are specifically built for finding the optimal parameters. For all studied systems, it was typically seen that choosing $q \tau = O(\tau_d)$ leads to accurate results; the optimal $q$ however was not found to be necessarily equal to $\tau_d/\tau$, and in some cases it could be as low as one-fifth of this value. It should be noted that the decorrelation timescale $\tau_d$ provides a measure of dynamical system's memory. We also observed that the optimal hard threshold presented in \citet{Brunton2019} yields a good criterion for the truncation value $r$. We again highlight that the optimal value of $r$ might be somewhat lower than what is given by this hard thresholding, which might be associated with the shortage of data. Nonetheless, the results are not too sensitive to the choice of $r$, so long as the selected $r$ is not very far from the threshold. Choosing $r$ significantly larger than this criterion leads to a rapidly diverging model, since many eigenvalues of $\mrm{A}$ fall outside the unit circle in this case. Finally, it was always observed that taking $p = \alpha r$ results in the most accurate predictions, where $\alpha$ is the ratio of the length of forcing vector to the length of state vector.  



\bibliography{Main}

\begin{thebibliography}{56}%
\makeatletter
\providecommand \@ifxundefined [1]{%
 \@ifx{#1\undefined}
}%
\providecommand \@ifnum [1]{%
 \ifnum #1\expandafter \@firstoftwo
 \else \expandafter \@secondoftwo
 \fi
}%
\providecommand \@ifx [1]{%
 \ifx #1\expandafter \@firstoftwo
 \else \expandafter \@secondoftwo
 \fi
}%
\providecommand \natexlab [1]{#1}%
\providecommand \enquote  [1]{``#1''}%
\providecommand \bibnamefont  [1]{#1}%
\providecommand \bibfnamefont [1]{#1}%
\providecommand \citenamefont [1]{#1}%
\providecommand \href@noop [0]{\@secondoftwo}%
\providecommand \href [0]{\begingroup \@sanitize@url \@href}%
\providecommand \@href[1]{\@@startlink{#1}\@@href}%
\providecommand \@@href[1]{\endgroup#1\@@endlink}%
\providecommand \@sanitize@url [0]{\catcode `\\12\catcode `\$12\catcode
  `\&12\catcode `\#12\catcode `\^12\catcode `\_12\catcode `\%12\relax}%
\providecommand \@@startlink[1]{}%
\providecommand \@@endlink[0]{}%
\providecommand \url  [0]{\begingroup\@sanitize@url \@url }%
\providecommand \@url [1]{\endgroup\@href {#1}{\urlprefix }}%
\providecommand \urlprefix  [0]{URL }%
\providecommand \Eprint [0]{\href }%
\providecommand \doibase [0]{http://dx.doi.org/}%
\providecommand \selectlanguage [0]{\@gobble}%
\providecommand \bibinfo  [0]{\@secondoftwo}%
\providecommand \bibfield  [0]{\@secondoftwo}%
\providecommand \translation [1]{[#1]}%
\providecommand \BibitemOpen [0]{}%
\providecommand \bibitemStop [0]{}%
\providecommand \bibitemNoStop [0]{.\EOS\space}%
\providecommand \EOS [0]{\spacefactor3000\relax}%
\providecommand \BibitemShut  [1]{\csname bibitem#1\endcsname}%
\let\auto@bib@innerbib\@empty
\bibitem [{\citenamefont {Lorenz}(1963)}]{Lorenz1963}%
  \BibitemOpen
  \bibfield  {author} {\bibinfo {author} {\bibfnamefont {E.~N.}\ \bibnamefont
  {Lorenz}},\ }\bibfield  {title} {\enquote {\bibinfo {title} {Deterministic
  nonperiodic flow},}\ }\href@noop {} {\bibfield  {journal} {\bibinfo
  {journal} {J. Atoms. Sci.}\ }\textbf {\bibinfo {volume} {20}},\ \bibinfo
  {pages} {130--141} (\bibinfo {year} {1963})}\BibitemShut {NoStop}%
\bibitem [{\citenamefont {Box}\ \emph {et~al.}(2015)\citenamefont {Box},
  \citenamefont {M.}, \citenamefont {Reinsel},\ and\ \citenamefont
  {Ljung}}]{Box2015}%
  \BibitemOpen
  \bibfield  {author} {\bibinfo {author} {\bibfnamefont {G.~E.~P.}\
  \bibnamefont {Box}}, \bibinfo {author} {\bibfnamefont {Jenkins~G.}\
  \bibnamefont {M.}}, \bibinfo {author} {\bibfnamefont {G.~C.}\ \bibnamefont
  {Reinsel}}, \ and\ \bibinfo {author} {\bibfnamefont {G.~M.}\ \bibnamefont
  {Ljung}},\ }\href@noop {} {\emph {\bibinfo {title} {Time series analysis:
  forecasting and control}}}\ (\bibinfo  {publisher} {John Wiley \& Sons},\
  \bibinfo {year} {2015})\BibitemShut {NoStop}%
\bibitem [{\citenamefont {Van~Kuik}\ \emph {et~al.}(2016)\citenamefont
  {Van~Kuik} \emph {et~al.}}]{van2016long}%
  \BibitemOpen
  \bibfield  {author} {\bibinfo {author} {\bibfnamefont {G.}~\bibnamefont
  {Van~Kuik}} \emph {et~al.},\ }\bibfield  {title} {\enquote {\bibinfo {title}
  {Long-term research challenges in wind energy - {A} research agenda by the
  {E}uropean academy of wind energy},}\ }\href@noop {} {\bibfield  {journal}
  {\bibinfo  {journal} {Wind Energ. Sci.}\ }\textbf {\bibinfo {volume} {1}},\
  \bibinfo {pages} {1--39} (\bibinfo {year} {2016})}\BibitemShut {NoStop}%
\bibitem [{\citenamefont {Duriez}\ \emph {et~al.}(2017)\citenamefont {Duriez},
  \citenamefont {Brunton},\ and\ \citenamefont {Noack}}]{duriez2017machine}%
  \BibitemOpen
  \bibfield  {author} {\bibinfo {author} {\bibfnamefont {T.}~\bibnamefont
  {Duriez}}, \bibinfo {author} {\bibfnamefont {S.~L.}\ \bibnamefont {Brunton}},
  \ and\ \bibinfo {author} {\bibfnamefont {B.~R.}\ \bibnamefont {Noack}},\
  }\href@noop {} {\emph {\bibinfo {title} {Machine Learning Control-Taming
  Nonlinear Dynamics and Turbulence}}}\ (\bibinfo  {publisher} {Springer},\
  \bibinfo {year} {2017})\BibitemShut {NoStop}%
\bibitem [{\citenamefont {Majda}(2012)}]{majda2012challenges}%
  \BibitemOpen
  \bibfield  {author} {\bibinfo {author} {\bibfnamefont {A.~J.}\ \bibnamefont
  {Majda}},\ }\bibfield  {title} {\enquote {\bibinfo {title} {Challenges in
  climate science and contemporary applied mathematics},}\ }\href@noop {}
  {\bibfield  {journal} {\bibinfo  {journal} {Commun. Pure Appl. Math.}\
  }\textbf {\bibinfo {volume} {65}},\ \bibinfo {pages} {920--948} (\bibinfo
  {year} {2012})}\BibitemShut {NoStop}%
\bibitem [{\citenamefont {Bauer}\ \emph {et~al.}(2015)\citenamefont {Bauer},
  \citenamefont {Thorpe},\ and\ \citenamefont {Brunet}}]{bauer2015quiet}%
  \BibitemOpen
  \bibfield  {author} {\bibinfo {author} {\bibfnamefont {P.}~\bibnamefont
  {Bauer}}, \bibinfo {author} {\bibfnamefont {A.}~\bibnamefont {Thorpe}}, \
  and\ \bibinfo {author} {\bibfnamefont {G.}~\bibnamefont {Brunet}},\
  }\bibfield  {title} {\enquote {\bibinfo {title} {The quiet revolution of
  numerical weather prediction},}\ }\href@noop {} {\bibfield  {journal}
  {\bibinfo  {journal} {Nature}\ }\textbf {\bibinfo {volume} {525}},\ \bibinfo
  {pages} {47} (\bibinfo {year} {2015})}\BibitemShut {NoStop}%
\bibitem [{\citenamefont {Majda}\ and\ \citenamefont
  {Chen}(2018)}]{majda2018model}%
  \BibitemOpen
  \bibfield  {author} {\bibinfo {author} {\bibfnamefont {A.}~\bibnamefont
  {Majda}}\ and\ \bibinfo {author} {\bibfnamefont {N.}~\bibnamefont {Chen}},\
  }\bibfield  {title} {\enquote {\bibinfo {title} {Model error, information
  barriers, state estimation and prediction in complex multiscale systems},}\
  }\href@noop {} {\bibfield  {journal} {\bibinfo  {journal} {Entropy}\ }\textbf
  {\bibinfo {volume} {20}},\ \bibinfo {pages} {644} (\bibinfo {year}
  {2018})}\BibitemShut {NoStop}%
\bibitem [{\citenamefont {Farazmand}\ and\ \citenamefont
  {Sapsis}(2017)}]{farazmand2017variational}%
  \BibitemOpen
  \bibfield  {author} {\bibinfo {author} {\bibfnamefont {M.}~\bibnamefont
  {Farazmand}}\ and\ \bibinfo {author} {\bibfnamefont {T.~P.}\ \bibnamefont
  {Sapsis}},\ }\bibfield  {title} {\enquote {\bibinfo {title} {A variational
  approach to probing extreme events in turbulent dynamical systems},}\
  }\href@noop {} {\bibfield  {journal} {\bibinfo  {journal} {Sci. Adv.}\
  }\textbf {\bibinfo {volume} {3}} (\bibinfo {year} {2017})}\BibitemShut
  {NoStop}%
\bibitem [{\citenamefont {Wunsch}(1999)}]{wunsch1999interpretation}%
  \BibitemOpen
  \bibfield  {author} {\bibinfo {author} {\bibfnamefont {C.}~\bibnamefont
  {Wunsch}},\ }\bibfield  {title} {\enquote {\bibinfo {title} {The
  interpretation of short climate records, with comments on the {North Atlantic
  and Southern Oscillations}},}\ }\href@noop {} {\bibfield  {journal} {\bibinfo
   {journal} {Bull. Am. Meteorol. Soc.}\ }\textbf {\bibinfo {volume} {80}},\
  \bibinfo {pages} {245--256} (\bibinfo {year} {1999})}\BibitemShut {NoStop}%
\bibitem [{\citenamefont {Van~den Dool}(2007)}]{van2007empirical}%
  \BibitemOpen
  \bibfield  {author} {\bibinfo {author} {\bibfnamefont {H.}~\bibnamefont
  {Van~den Dool}},\ }\href@noop {} {\emph {\bibinfo {title} {Empirical methods
  in short-term climate prediction}}}\ (\bibinfo  {publisher} {Oxford
  University Press},\ \bibinfo {year} {2007})\BibitemShut {NoStop}%
\bibitem [{\citenamefont {Cavanaugh}\ \emph {et~al.}(2015)\citenamefont
  {Cavanaugh}, \citenamefont {Allen}, \citenamefont {Subramanian},
  \citenamefont {Mapes}, \citenamefont {Seo},\ and\ \citenamefont
  {Miller}}]{cavanaugh2015skill}%
  \BibitemOpen
  \bibfield  {author} {\bibinfo {author} {\bibfnamefont {N.~R.}\ \bibnamefont
  {Cavanaugh}}, \bibinfo {author} {\bibfnamefont {T.}~\bibnamefont {Allen}},
  \bibinfo {author} {\bibfnamefont {A.}~\bibnamefont {Subramanian}}, \bibinfo
  {author} {\bibfnamefont {B.}~\bibnamefont {Mapes}}, \bibinfo {author}
  {\bibfnamefont {H.}~\bibnamefont {Seo}}, \ and\ \bibinfo {author}
  {\bibfnamefont {A.~J.}\ \bibnamefont {Miller}},\ }\bibfield  {title}
  {\enquote {\bibinfo {title} {The skill of atmospheric linear inverse models
  in hindcasting the {Madden--Julian O}scillation},}\ }\href@noop {} {\bibfield
   {journal} {\bibinfo  {journal} {Clim. Dyn.}\ }\textbf {\bibinfo {volume}
  {44}},\ \bibinfo {pages} {897--906} (\bibinfo {year} {2015})}\BibitemShut
  {NoStop}%
\bibitem [{\citenamefont {Giannakis}(2017)}]{Giannakis2017}%
  \BibitemOpen
  \bibfield  {author} {\bibinfo {author} {\bibfnamefont {D.}~\bibnamefont
  {Giannakis}},\ }\bibfield  {title} {\enquote {\bibinfo {title} {Data-driven
  spectral decomposition and forecasting of ergodic dynamical systems},}\
  }\href@noop {} {\bibfield  {journal} {\bibinfo  {journal} {Appl. Comput.
  Harmon. Anal.}\ } (\bibinfo {year} {2017})}\BibitemShut {NoStop}%
\bibitem [{\citenamefont {Comeau}\ \emph {et~al.}(2017)\citenamefont {Comeau},
  \citenamefont {Zhao}, \citenamefont {Giannakis},\ and\ \citenamefont
  {Majda}}]{comeau2017data}%
  \BibitemOpen
  \bibfield  {author} {\bibinfo {author} {\bibfnamefont {D.}~\bibnamefont
  {Comeau}}, \bibinfo {author} {\bibfnamefont {Z.}~\bibnamefont {Zhao}},
  \bibinfo {author} {\bibfnamefont {D.}~\bibnamefont {Giannakis}}, \ and\
  \bibinfo {author} {\bibfnamefont {A.~J.}\ \bibnamefont {Majda}},\ }\bibfield
  {title} {\enquote {\bibinfo {title} {Data-driven prediction strategies for
  low-frequency patterns of {North P}acific climate variability},}\ }\href@noop
  {} {\bibfield  {journal} {\bibinfo  {journal} {Clima. Dyn.}\ }\textbf
  {\bibinfo {volume} {48}},\ \bibinfo {pages} {1855--1872} (\bibinfo {year}
  {2017})}\BibitemShut {NoStop}%
\bibitem [{\citenamefont {Khodkar}\ and\ \citenamefont
  {Hassanzadeh}(2018)}]{Khodkar2018}%
  \BibitemOpen
  \bibfield  {author} {\bibinfo {author} {\bibfnamefont {M.~A.}\ \bibnamefont
  {Khodkar}}\ and\ \bibinfo {author} {\bibfnamefont {P.}~\bibnamefont
  {Hassanzadeh}},\ }\bibfield  {title} {\enquote {\bibinfo {title} {Data-driven
  reduced modelling of turbulent {R}ayleigh-{B}\'{e}nard convection using
  dmd-enhanced fluctuation-dissipation theorem},}\ }\href {\doibase
  10.1017/jfm.2018.586} {\bibfield  {journal} {\bibinfo  {journal} {J. Fluid
  Mech.}\ }\textbf {\bibinfo {volume} {852}} (\bibinfo {year} {2018}),\
  10.1017/jfm.2018.586}\BibitemShut {NoStop}%
\bibitem [{\citenamefont {McDermott}\ and\ \citenamefont
  {Wikle}(2017)}]{McDermott2017}%
  \BibitemOpen
  \bibfield  {author} {\bibinfo {author} {\bibfnamefont {P.~L.}\ \bibnamefont
  {McDermott}}\ and\ \bibinfo {author} {\bibfnamefont {C.~K.}\ \bibnamefont
  {Wikle}},\ }\bibfield  {title} {\enquote {\bibinfo {title} {An ensemble
  quadratic echo state network for non-linear spatio-temporal forecasting},}\
  }\href@noop {} {\bibfield  {journal} {\bibinfo  {journal} {Stat}\ }\textbf
  {\bibinfo {volume} {6}},\ \bibinfo {pages} {315--330} (\bibinfo {year}
  {2017})}\BibitemShut {NoStop}%
\bibitem [{\citenamefont {Yu}\ \emph {et~al.}(2017)\citenamefont {Yu},
  \citenamefont {Zheng}, \citenamefont {Anandkumar},\ and\ \citenamefont
  {Yue}}]{yu2017long}%
  \BibitemOpen
  \bibfield  {author} {\bibinfo {author} {\bibfnamefont {R.}~\bibnamefont
  {Yu}}, \bibinfo {author} {\bibfnamefont {S.}~\bibnamefont {Zheng}}, \bibinfo
  {author} {\bibfnamefont {A.}~\bibnamefont {Anandkumar}}, \ and\ \bibinfo
  {author} {\bibfnamefont {Y.}~\bibnamefont {Yue}},\ }\bibfield  {title}
  {\enquote {\bibinfo {title} {Long-term forecasting using tensor-train
  {RNN}s},}\ }\href@noop {} {\bibfield  {journal} {\bibinfo  {journal}
  {arXiv:1711.00073}\ } (\bibinfo {year} {2017})}\BibitemShut {NoStop}%
\bibitem [{\citenamefont {Vlachas}\ \emph {et~al.}(2018)\citenamefont
  {Vlachas}, \citenamefont {Byeon}, \citenamefont {Wan}, \citenamefont
  {Sapsis},\ and\ \citenamefont {Koumoutsakos}}]{Vlachas2018}%
  \BibitemOpen
  \bibfield  {author} {\bibinfo {author} {\bibfnamefont {P.~R.}\ \bibnamefont
  {Vlachas}}, \bibinfo {author} {\bibfnamefont {W.}~\bibnamefont {Byeon}},
  \bibinfo {author} {\bibfnamefont {Z.~Y.}\ \bibnamefont {Wan}}, \bibinfo
  {author} {\bibfnamefont {T.}~\bibnamefont {Sapsis}}, \ and\ \bibinfo {author}
  {\bibfnamefont {P.}~\bibnamefont {Koumoutsakos}},\ }\bibfield  {title}
  {\enquote {\bibinfo {title} {Data-driven forecasting of high-dimensional
  chaotic systems with long short-term memory networks},}\ }\href@noop {}
  {\bibfield  {journal} {\bibinfo  {journal} {Proc. R. Soc. A}\ }\textbf
  {\bibinfo {volume} {474}} (\bibinfo {year} {2018})}\BibitemShut {NoStop}%
\bibitem [{\citenamefont {Pathak}\ \emph {et~al.}(2018)\citenamefont {Pathak},
  \citenamefont {Hunt}, \citenamefont {Girvan}, \citenamefont {Lu},\ and\
  \citenamefont {Ott}}]{Pathak2018}%
  \BibitemOpen
  \bibfield  {author} {\bibinfo {author} {\bibfnamefont {U.}~\bibnamefont
  {Pathak}}, \bibinfo {author} {\bibfnamefont {B.}~\bibnamefont {Hunt}},
  \bibinfo {author} {\bibfnamefont {M.}~\bibnamefont {Girvan}}, \bibinfo
  {author} {\bibfnamefont {Z.}~\bibnamefont {Lu}}, \ and\ \bibinfo {author}
  {\bibfnamefont {E.}~\bibnamefont {Ott}},\ }\bibfield  {title} {\enquote
  {\bibinfo {title} {Model-free prediction of large spatiotemporally chaotic
  systems from data: {A} reservoir computing approach},}\ }\href@noop {}
  {\bibfield  {journal} {\bibinfo  {journal} {Phys. Rev. Lett.}\ }\textbf
  {\bibinfo {volume} {120}} (\bibinfo {year} {2018})}\BibitemShut {NoStop}%
\bibitem [{\citenamefont {Raissi}\ \emph {et~al.}(2019)\citenamefont {Raissi},
  \citenamefont {Perdikaris},\ and\ \citenamefont {Karniadakis}}]{Raissi2019}%
  \BibitemOpen
  \bibfield  {author} {\bibinfo {author} {\bibfnamefont {M.}~\bibnamefont
  {Raissi}}, \bibinfo {author} {\bibfnamefont {P.}~\bibnamefont {Perdikaris}},
  \ and\ \bibinfo {author} {\bibfnamefont {G.~E.}\ \bibnamefont
  {Karniadakis}},\ }\bibfield  {title} {\enquote {\bibinfo {title}
  {Physics-informed neural networks: {A} deep learning framework for learning
  forward and inverse problems involving nonlinear partial differential
  equations},}\ }\href@noop {} {\bibfield  {journal} {\bibinfo  {journal} {J.
  Comp. Phys.}\ }\textbf {\bibinfo {volume} {378}},\ \bibinfo {pages}
  {686--707} (\bibinfo {year} {2019})}\BibitemShut {NoStop}%
\bibitem [{\citenamefont {Mohan}\ \emph {et~al.}(2019)\citenamefont {Mohan},
  \citenamefont {Daniel}, \citenamefont {Chertkov},\ and\ \citenamefont
  {Livescu}}]{Mohan2019}%
  \BibitemOpen
  \bibfield  {author} {\bibinfo {author} {\bibfnamefont {A.~T.}\ \bibnamefont
  {Mohan}}, \bibinfo {author} {\bibfnamefont {D.}~\bibnamefont {Daniel}},
  \bibinfo {author} {\bibfnamefont {M.}~\bibnamefont {Chertkov}}, \ and\
  \bibinfo {author} {\bibfnamefont {D.}~\bibnamefont {Livescu}},\ }\bibfield
  {title} {\enquote {\bibinfo {title} {Compressed convolutional {LSTM}: {A}n
  efficient deep learning framework to model high fidelity 3{D} turbulence},}\
  }\href@noop {} {\bibfield  {journal} {\bibinfo  {journal} {arXiv:1903.00033}\
  } (\bibinfo {year} {2019})}\BibitemShut {NoStop}%
\bibitem [{\citenamefont {McDermott}\ and\ \citenamefont
  {Wikle}(2019)}]{McDermott2019}%
  \BibitemOpen
  \bibfield  {author} {\bibinfo {author} {\bibfnamefont {P.~L.}\ \bibnamefont
  {McDermott}}\ and\ \bibinfo {author} {\bibfnamefont {C.~K.}\ \bibnamefont
  {Wikle}},\ }\bibfield  {title} {\enquote {\bibinfo {title} {Deep state
  networks with uncertainty quantification for spatio-temporal forecasting},}\
  }\href@noop {} {\bibfield  {journal} {\bibinfo  {journal} {Environmetrics}\
  }\textbf {\bibinfo {volume} {30}},\ \bibinfo {pages} {e2553} (\bibinfo {year}
  {2019})}\BibitemShut {NoStop}%
\bibitem [{\citenamefont {Chattopadhyay}\ \emph {et~al.}(2019)\citenamefont
  {Chattopadhyay}, \citenamefont {Hassanzadeh}, \citenamefont {Palem},\ and\
  \citenamefont {Subramanian}}]{Ashesh2019}%
  \BibitemOpen
  \bibfield  {author} {\bibinfo {author} {\bibfnamefont {A.}~\bibnamefont
  {Chattopadhyay}}, \bibinfo {author} {\bibfnamefont {P.}~\bibnamefont
  {Hassanzadeh}}, \bibinfo {author} {\bibfnamefont {K.}~\bibnamefont {Palem}},
  \ and\ \bibinfo {author} {\bibfnamefont {D.}~\bibnamefont {Subramanian}},\
  }\bibfield  {title} {\enquote {\bibinfo {title} {Data-driven prediction of a
  multi-scale lorenz96 chaotic system using a hierarchy of deep learning
  methods: {R}eservoir computing, {ANN}, and {RNN}-{LSTM}},}\ }\href@noop {}
  {\bibfield  {journal} {\bibinfo  {journal} {arXiv:1906.08829}\ } (\bibinfo
  {year} {2019})}\BibitemShut {NoStop}%
\bibitem [{\citenamefont {Koopman}(1931)}]{koopman1931hamiltonian}%
  \BibitemOpen
  \bibfield  {author} {\bibinfo {author} {\bibfnamefont {B.~O.}\ \bibnamefont
  {Koopman}},\ }\bibfield  {title} {\enquote {\bibinfo {title} {Hamiltonian
  systems and transformation in {H}ilbert space},}\ }\href@noop {} {\bibfield
  {journal} {\bibinfo  {journal} {Proc. Natl. Acad. Sci.}\ }\textbf {\bibinfo
  {volume} {17}} (\bibinfo {year} {1931})}\BibitemShut {NoStop}%
\bibitem [{\citenamefont {Mezi{\'c}}(2005)}]{mezic2005spectral}%
  \BibitemOpen
  \bibfield  {author} {\bibinfo {author} {\bibfnamefont {I.}~\bibnamefont
  {Mezi{\'c}}},\ }\bibfield  {title} {\enquote {\bibinfo {title} {Spectral
  properties of dynamical systems, model reduction and decompositions},}\
  }\href@noop {} {\bibfield  {journal} {\bibinfo  {journal} {Nonlin. Dyn.}\
  }\textbf {\bibinfo {volume} {41}},\ \bibinfo {pages} {309--325} (\bibinfo
  {year} {2005})}\BibitemShut {NoStop}%
\bibitem [{\citenamefont {Mezi{\'c}}(2013)}]{mezic2013analysis}%
  \BibitemOpen
  \bibfield  {author} {\bibinfo {author} {\bibfnamefont {I.}~\bibnamefont
  {Mezi{\'c}}},\ }\bibfield  {title} {\enquote {\bibinfo {title} {Analysis of
  fluid flows via spectral properties of the {K}oopman operator},}\ }\href@noop
  {} {\bibfield  {journal} {\bibinfo  {journal} {Annu. Rev. Fluid Mech.}\
  }\textbf {\bibinfo {volume} {45}} (\bibinfo {year} {2013})}\BibitemShut
  {NoStop}%
\bibitem [{\citenamefont {Schmid}(2010)}]{Schmid2010}%
  \BibitemOpen
  \bibfield  {author} {\bibinfo {author} {\bibfnamefont {P.~J.}\ \bibnamefont
  {Schmid}},\ }\bibfield  {title} {\enquote {\bibinfo {title} {Dynamic mode
  decomposition of numerical and experimental data},}\ }\href@noop {}
  {\bibfield  {journal} {\bibinfo  {journal} {J. Fluid Mech.}\ }\textbf
  {\bibinfo {volume} {656}},\ \bibinfo {pages} {5--28} (\bibinfo {year}
  {2010})}\BibitemShut {NoStop}%
\bibitem [{\citenamefont {Rowley}\ \emph {et~al.}(2009)\citenamefont {Rowley},
  \citenamefont {Mezi{\'c}}, \citenamefont {Bagheri}, \citenamefont
  {Schlatter},\ and\ \citenamefont {Henningson}}]{rowley2009spectral}%
  \BibitemOpen
  \bibfield  {author} {\bibinfo {author} {\bibfnamefont {C.~W.}\ \bibnamefont
  {Rowley}}, \bibinfo {author} {\bibfnamefont {I.}~\bibnamefont {Mezi{\'c}}},
  \bibinfo {author} {\bibfnamefont {S.}~\bibnamefont {Bagheri}}, \bibinfo
  {author} {\bibfnamefont {P.}~\bibnamefont {Schlatter}}, \ and\ \bibinfo
  {author} {\bibfnamefont {D.~S.}\ \bibnamefont {Henningson}},\ }\bibfield
  {title} {\enquote {\bibinfo {title} {Spectral analysis of nonlinear flows},}\
  }\href@noop {} {\bibfield  {journal} {\bibinfo  {journal} {J. Fluid Mech.}\
  }\textbf {\bibinfo {volume} {641}},\ \bibinfo {pages} {115--127} (\bibinfo
  {year} {2009})}\BibitemShut {NoStop}%
\bibitem [{\citenamefont {Tu}\ \emph {et~al.}(2014)\citenamefont {Tu},
  \citenamefont {Rowley}, \citenamefont {M.}, \citenamefont {Brunton},\ and\
  \citenamefont {Kutz}}]{Tu2014}%
  \BibitemOpen
  \bibfield  {author} {\bibinfo {author} {\bibfnamefont {J.~H.}\ \bibnamefont
  {Tu}}, \bibinfo {author} {\bibfnamefont {C.~W.}\ \bibnamefont {Rowley}},
  \bibinfo {author} {\bibfnamefont {Luchtenburg~D.}\ \bibnamefont {M.}},
  \bibinfo {author} {\bibfnamefont {S.~L.}\ \bibnamefont {Brunton}}, \ and\
  \bibinfo {author} {\bibfnamefont {J.~N.}\ \bibnamefont {Kutz}},\ }\bibfield
  {title} {\enquote {\bibinfo {title} {On dynamic mode decomposition: {T}heory
  and applications},}\ }\href@noop {} {\bibfield  {journal} {\bibinfo
  {journal} {J. Comp. Dyn.}\ }\textbf {\bibinfo {volume} {1}},\ \bibinfo
  {pages} {391--421} (\bibinfo {year} {2014})}\BibitemShut {NoStop}%
\bibitem [{\citenamefont {Williams}\ \emph {et~al.}(2015)\citenamefont
  {Williams}, \citenamefont {Kevrekidis},\ and\ \citenamefont
  {Rowley}}]{williams2015data}%
  \BibitemOpen
  \bibfield  {author} {\bibinfo {author} {\bibfnamefont {M.~O.}\ \bibnamefont
  {Williams}}, \bibinfo {author} {\bibfnamefont {I.~G.}\ \bibnamefont
  {Kevrekidis}}, \ and\ \bibinfo {author} {\bibfnamefont {C.~W.}\ \bibnamefont
  {Rowley}},\ }\bibfield  {title} {\enquote {\bibinfo {title} {A data--driven
  approximation of the {K}oopman operator: Extending dynamic mode
  decomposition},}\ }\href@noop {} {\bibfield  {journal} {\bibinfo  {journal}
  {J. Nonlin. Sci.}\ }\textbf {\bibinfo {volume} {25}} (\bibinfo {year}
  {2015})}\BibitemShut {NoStop}%
\bibitem [{\citenamefont {Arbabi}\ and\ \citenamefont
  {Mezi\'{c}}(2017)}]{Arbabi2017}%
  \BibitemOpen
  \bibfield  {author} {\bibinfo {author} {\bibfnamefont {H.}~\bibnamefont
  {Arbabi}}\ and\ \bibinfo {author} {\bibfnamefont {I.}~\bibnamefont
  {Mezi\'{c}}},\ }\bibfield  {title} {\enquote {\bibinfo {title} {Ergodic
  theory, dynamic mode decomposition, and computation of spectral properties of
  the {K}oopman operator},}\ }\href@noop {} {\bibfield  {journal} {\bibinfo
  {journal} {SIAM J. Appl. Dyn. Syst.}\ }\textbf {\bibinfo {volume} {16}},\
  \bibinfo {pages} {2096--2126} (\bibinfo {year} {2017})}\BibitemShut {NoStop}%
\bibitem [{\citenamefont {Arbabi}\ and\ \citenamefont
  {Mezi{\'c}}(2017)}]{Arbabi2017study}%
  \BibitemOpen
  \bibfield  {author} {\bibinfo {author} {\bibfnamefont {H.}~\bibnamefont
  {Arbabi}}\ and\ \bibinfo {author} {\bibfnamefont {I.}~\bibnamefont
  {Mezi{\'c}}},\ }\bibfield  {title} {\enquote {\bibinfo {title} {Study of
  dynamics in post-transient flows using {K}oopman mode decomposition},}\
  }\href@noop {} {\bibfield  {journal} {\bibinfo  {journal} {Phys. Rev.
  Fluids}\ }\textbf {\bibinfo {volume} {2}},\ \bibinfo {pages} {124402}
  (\bibinfo {year} {2017})}\BibitemShut {NoStop}%
\bibitem [{\citenamefont {Korda}\ and\ \citenamefont
  {Mezi{\'c}}(2018{\natexlab{a}})}]{korda2018convergence}%
  \BibitemOpen
  \bibfield  {author} {\bibinfo {author} {\bibfnamefont {M.}~\bibnamefont
  {Korda}}\ and\ \bibinfo {author} {\bibfnamefont {I.}~\bibnamefont
  {Mezi{\'c}}},\ }\bibfield  {title} {\enquote {\bibinfo {title} {On
  convergence of extended dynamic mode decomposition to the {K}oopman
  operator},}\ }\href@noop {} {\bibfield  {journal} {\bibinfo  {journal} {J.
  Nonlin. Sci.}\ }\textbf {\bibinfo {volume} {28}},\ \bibinfo {pages}
  {687--710} (\bibinfo {year} {2018}{\natexlab{a}})}\BibitemShut {NoStop}%
\bibitem [{\citenamefont {Rowley}\ and\ \citenamefont
  {Dawson}(2017)}]{Rowley2017}%
  \BibitemOpen
  \bibfield  {author} {\bibinfo {author} {\bibfnamefont {C.~W.}\ \bibnamefont
  {Rowley}}\ and\ \bibinfo {author} {\bibfnamefont {S.~T.~M.}\ \bibnamefont
  {Dawson}},\ }\bibfield  {title} {\enquote {\bibinfo {title} {Model reduction
  for flow analysis and control},}\ }\href@noop {} {\bibfield  {journal}
  {\bibinfo  {journal} {Annu. Rev. Fluid Mech.}\ }\textbf {\bibinfo {volume}
  {49}},\ \bibinfo {pages} {387--417} (\bibinfo {year} {2017})}\BibitemShut
  {NoStop}%
\bibitem [{\citenamefont {Takens}(1981)}]{Takens1981}%
  \BibitemOpen
  \bibfield  {author} {\bibinfo {author} {\bibfnamefont {F.}~\bibnamefont
  {Takens}},\ }\bibfield  {title} {\enquote {\bibinfo {title} {Detecting
  strange attractors in turbulence},}\ }\href@noop {} {\bibfield  {journal}
  {\bibinfo  {journal} {Lect. Notes Math.}\ }\textbf {\bibinfo {volume}
  {898}},\ \bibinfo {pages} {366--381} (\bibinfo {year} {1981})}\BibitemShut
  {NoStop}%
\bibitem [{\citenamefont {Korda}\ and\ \citenamefont
  {Mezi{\'c}}(2018{\natexlab{b}})}]{Korda2018}%
  \BibitemOpen
  \bibfield  {author} {\bibinfo {author} {\bibfnamefont {M.}~\bibnamefont
  {Korda}}\ and\ \bibinfo {author} {\bibfnamefont {I.}~\bibnamefont
  {Mezi{\'c}}},\ }\bibfield  {title} {\enquote {\bibinfo {title} {Linear
  predictors for nonlinear dynamical systems: {K}oopman operator meets model
  predictive control},}\ }\href@noop {} {\bibfield  {journal} {\bibinfo
  {journal} {Automatica}\ }\textbf {\bibinfo {volume} {93}},\ \bibinfo {pages}
  {149--160} (\bibinfo {year} {2018}{\natexlab{b}})}\BibitemShut {NoStop}%
\bibitem [{\citenamefont {Arbabi}\ \emph {et~al.}(2018)\citenamefont {Arbabi},
  \citenamefont {Korda},\ and\ \citenamefont {Mezi\'{c}}}]{Arbabi2018}%
  \BibitemOpen
  \bibfield  {author} {\bibinfo {author} {\bibfnamefont {H.}~\bibnamefont
  {Arbabi}}, \bibinfo {author} {\bibfnamefont {M.}~\bibnamefont {Korda}}, \
  and\ \bibinfo {author} {\bibfnamefont {I.}~\bibnamefont {Mezi\'{c}}},\
  }\bibfield  {title} {\enquote {\bibinfo {title} {A data-driven {K}oopman
  model predictive control for nonlinear flows},}\ }\href@noop {} {\bibfield
  {journal} {\bibinfo  {journal} {arXiv:1804.05291}\ } (\bibinfo {year}
  {2018})}\BibitemShut {NoStop}%
\bibitem [{\citenamefont {Giannakis}\ and\ \citenamefont
  {Majda}(2012)}]{Giannakis2012}%
  \BibitemOpen
  \bibfield  {author} {\bibinfo {author} {\bibfnamefont {D.}~\bibnamefont
  {Giannakis}}\ and\ \bibinfo {author} {\bibfnamefont {A.~J.}\ \bibnamefont
  {Majda}},\ }\bibfield  {title} {\enquote {\bibinfo {title} {Nonlinear
  laplacian spectral analysis for time series with intermittency and
  low-frequency variability},}\ }\href@noop {} {\bibfield  {journal} {\bibinfo
  {journal} {Proc. Natl Acad. Sci.}\ }\textbf {\bibinfo {volume} {113}},\
  \bibinfo {pages} {3932--3937} (\bibinfo {year} {2012})}\BibitemShut {NoStop}%
\bibitem [{\citenamefont {Brunton}\ \emph {et~al.}(2017)\citenamefont
  {Brunton}, \citenamefont {Brunton}, \citenamefont {Proctor}, \citenamefont
  {Kaiser},\ and\ \citenamefont {Kutz}}]{Brunton2017}%
  \BibitemOpen
  \bibfield  {author} {\bibinfo {author} {\bibfnamefont {S.~L.}\ \bibnamefont
  {Brunton}}, \bibinfo {author} {\bibfnamefont {B.~W.}\ \bibnamefont
  {Brunton}}, \bibinfo {author} {\bibfnamefont {J.~L.}\ \bibnamefont
  {Proctor}}, \bibinfo {author} {\bibfnamefont {E.}~\bibnamefont {Kaiser}}, \
  and\ \bibinfo {author} {\bibfnamefont {J.~N.}\ \bibnamefont {Kutz}},\
  }\bibfield  {title} {\enquote {\bibinfo {title} {Chaos as an intermittently
  forced linear system},}\ }\href@noop {} {\bibfield  {journal} {\bibinfo
  {journal} {Nat. Commun.}\ }\textbf {\bibinfo {volume} {8}},\ \bibinfo {pages}
  {19} (\bibinfo {year} {2017})}\BibitemShut {NoStop}%
\bibitem [{\citenamefont {Brunton}\ \emph {et~al.}(2016)\citenamefont
  {Brunton}, \citenamefont {Proctor}, \citenamefont {Kaiser},\ and\
  \citenamefont {Kutz}}]{Brunton2016b}%
  \BibitemOpen
  \bibfield  {author} {\bibinfo {author} {\bibfnamefont {S.~L.}\ \bibnamefont
  {Brunton}}, \bibinfo {author} {\bibfnamefont {J.~L.}\ \bibnamefont
  {Proctor}}, \bibinfo {author} {\bibfnamefont {E.}~\bibnamefont {Kaiser}}, \
  and\ \bibinfo {author} {\bibfnamefont {J.~N.}\ \bibnamefont {Kutz}},\
  }\bibfield  {title} {\enquote {\bibinfo {title} {Discovering governing
  equations from data by sparse identification of nonlinear dynamical
  systems},}\ }\href@noop {} {\bibfield  {journal} {\bibinfo  {journal} {Proc.
  Natl Acad. Sci.}\ }\textbf {\bibinfo {volume} {113}},\ \bibinfo {pages}
  {3932--3937} (\bibinfo {year} {2016})}\BibitemShut {NoStop}%
\bibitem [{\citenamefont {Ionita}\ and\ \citenamefont
  {Antoulas}(2012)}]{Thanos2012}%
  \BibitemOpen
  \bibfield  {author} {\bibinfo {author} {\bibfnamefont {A.~C.}\ \bibnamefont
  {Ionita}}\ and\ \bibinfo {author} {\bibfnamefont {A.~C.}\ \bibnamefont
  {Antoulas}},\ }\bibfield  {title} {\enquote {\bibinfo {title} {Matrix pencils
  in time and frequency domain system identification},}\ }in\ \href@noop {}
  {\emph {\bibinfo {booktitle} {Developments in Control Theory: Towards Glocal
  Control}}},\ Vol.~\bibinfo {volume} {76},\ \bibinfo {editor} {edited by\
  \bibinfo {editor} {\bibfnamefont {L.}~\bibnamefont {Qiu}}, \bibinfo {editor}
  {\bibfnamefont {J.}~\bibnamefont {Chen}}, \bibinfo {editor} {\bibfnamefont
  {T.}~\bibnamefont {Iwasaki}}, \ and\ \bibinfo {editor} {\bibfnamefont
  {H.}~\bibnamefont {Fujioka}}}\ (\bibinfo  {publisher} {IET Control
  Engineering Series},\ \bibinfo {year} {2012})\ pp.\ \bibinfo {pages}
  {79--88}\BibitemShut {NoStop}%
\bibitem [{\citenamefont {Antoulas}\ \emph {et~al.}(2016)\citenamefont
  {Antoulas}, \citenamefont {Gosea},\ and\ \citenamefont
  {Ionita}}]{Thanos2016}%
  \BibitemOpen
  \bibfield  {author} {\bibinfo {author} {\bibfnamefont {A.~C.}\ \bibnamefont
  {Antoulas}}, \bibinfo {author} {\bibfnamefont {I.~V.}\ \bibnamefont {Gosea}},
  \ and\ \bibinfo {author} {\bibfnamefont {A.~C.}\ \bibnamefont {Ionita}},\
  }\bibfield  {title} {\enquote {\bibinfo {title} {Model reduction of bilinear
  systems in the {L}oewner framework},}\ }\href@noop {} {\bibfield  {journal}
  {\bibinfo  {journal} {SIAM J. Sci. Comput.}\ }\textbf {\bibinfo {volume}
  {38}},\ \bibinfo {pages} {B889--B916} (\bibinfo {year} {2016})}\BibitemShut
  {NoStop}%
\bibitem [{\citenamefont {Antoulas}\ \emph {et~al.}(2019)\citenamefont
  {Antoulas}, \citenamefont {Gosea},\ and\ \citenamefont
  {Heinkenschloss}}]{Thanos2019a}%
  \BibitemOpen
  \bibfield  {author} {\bibinfo {author} {\bibfnamefont {A.~C.}\ \bibnamefont
  {Antoulas}}, \bibinfo {author} {\bibfnamefont {I.~V.}\ \bibnamefont {Gosea}},
  \ and\ \bibinfo {author} {\bibfnamefont {M.}~\bibnamefont {Heinkenschloss}},\
  }\bibfield  {title} {\enquote {\bibinfo {title} {On the {L}oewner framework
  for model reduction of {B}urgers' equation},}\ }in\ \href@noop {} {\emph
  {\bibinfo {booktitle} {Active Flow and Combustion Control 2018}}}\ (\bibinfo
  {publisher} {Springer},\ \bibinfo {year} {2019})\ pp.\ \bibinfo {pages}
  {255--270}\BibitemShut {NoStop}%
\bibitem [{\citenamefont {Gugercin}\ \emph {et~al.}(2019)\citenamefont
  {Gugercin}, \citenamefont {Beattie},\ and\ \citenamefont
  {Antoulas}}]{Thanos2019b}%
  \BibitemOpen
  \bibfield  {author} {\bibinfo {author} {\bibfnamefont {S.}~\bibnamefont
  {Gugercin}}, \bibinfo {author} {\bibfnamefont {C.~A.}\ \bibnamefont
  {Beattie}}, \ and\ \bibinfo {author} {\bibfnamefont {A.~C.}\ \bibnamefont
  {Antoulas}},\ }\href@noop {} {\emph {\bibinfo {title} {Data-driven and
  interpolatory model reduction}}}\ (\bibinfo  {publisher} {Society for
  Industrial and Applied Mathematics},\ \bibinfo {year} {2019})\BibitemShut
  {NoStop}%
\bibitem [{\citenamefont {Pogorelyuk}\ and\ \citenamefont
  {Rowley}(2018)}]{Pogorelyuk2018}%
  \BibitemOpen
  \bibfield  {author} {\bibinfo {author} {\bibfnamefont {L.}~\bibnamefont
  {Pogorelyuk}}\ and\ \bibinfo {author} {\bibfnamefont {C.~W.}\ \bibnamefont
  {Rowley}},\ }\bibfield  {title} {\enquote {\bibinfo {title} {Clustering of
  series via dynamic mode decomposition and the matrix pencil method},}\
  }\href@noop {} {\bibfield  {journal} {\bibinfo  {journal} {arXiv:1802.09878}\
  } (\bibinfo {year} {2018})}\BibitemShut {NoStop}%
\bibitem [{\citenamefont {Proctor}\ \emph {et~al.}(2016)\citenamefont
  {Proctor}, \citenamefont {Brunton},\ and\ \citenamefont
  {Kutz}}]{Brunton2016a}%
  \BibitemOpen
  \bibfield  {author} {\bibinfo {author} {\bibfnamefont {J.~L}\ \bibnamefont
  {Proctor}}, \bibinfo {author} {\bibfnamefont {S.~L.}\ \bibnamefont
  {Brunton}}, \ and\ \bibinfo {author} {\bibfnamefont {J.~N.}\ \bibnamefont
  {Kutz}},\ }\bibfield  {title} {\enquote {\bibinfo {title} {Dynamic mode
  decomposition with control},}\ }\href@noop {} {\bibfield  {journal} {\bibinfo
   {journal} {SIAM J. Appl. Dynam. Syst.}\ }\textbf {\bibinfo {volume} {15}},\
  \bibinfo {pages} {142--161} (\bibinfo {year} {2016})}\BibitemShut {NoStop}%
\bibitem [{\citenamefont {Wolf}\ \emph {et~al.}(1985)\citenamefont {Wolf},
  \citenamefont {Swift}, \citenamefont {Swinney},\ and\ \citenamefont
  {Vastano}}]{Wolf1985}%
  \BibitemOpen
  \bibfield  {author} {\bibinfo {author} {\bibfnamefont {A.}~\bibnamefont
  {Wolf}}, \bibinfo {author} {\bibfnamefont {J.~B.}\ \bibnamefont {Swift}},
  \bibinfo {author} {\bibfnamefont {H.~L.}\ \bibnamefont {Swinney}}, \ and\
  \bibinfo {author} {\bibfnamefont {J.~A.}\ \bibnamefont {Vastano}},\
  }\bibfield  {title} {\enquote {\bibinfo {title} {Determining {L}yapunov
  exponents from a time series},}\ }\href@noop {} {\bibfield  {journal}
  {\bibinfo  {journal} {Physica D}\ }\textbf {\bibinfo {volume} {16}},\
  \bibinfo {pages} {285--317} (\bibinfo {year} {1985})}\BibitemShut {NoStop}%
\bibitem [{\citenamefont {Sapsis}\ and\ \citenamefont
  {Majda}(2013)}]{Sapsis2013}%
  \BibitemOpen
  \bibfield  {author} {\bibinfo {author} {\bibfnamefont {T.}~\bibnamefont
  {Sapsis}}\ and\ \bibinfo {author} {\bibfnamefont {A.~J.}\ \bibnamefont
  {Majda}},\ }\bibfield  {title} {\enquote {\bibinfo {title} {Statistically
  accurate low-order models for uncertainty quantification in turbulent
  dynamical systems},}\ }\href@noop {} {\bibfield  {journal} {\bibinfo
  {journal} {Proc. Natl. Acad. Sci.}\ }\textbf {\bibinfo {volume} {110}}
  (\bibinfo {year} {2013})}\BibitemShut {NoStop}%
\bibitem [{\citenamefont {Majda}(2016)}]{Majda2016}%
  \BibitemOpen
  \bibfield  {author} {\bibinfo {author} {\bibfnamefont {A.~J.}\ \bibnamefont
  {Majda}},\ }\bibfield  {title} {\enquote {\bibinfo {title} {Introduction to
  turbulent dynamical systems for complex systems},}\ }in\ \href@noop {} {\emph
  {\bibinfo {booktitle} {Frontiers in Applied Dynamical Systems: Reviews and
  Tutorials}}}\ (\bibinfo  {publisher} {Springer},\ \bibinfo {year}
  {2016})\BibitemShut {NoStop}%
\bibitem [{\citenamefont {Qi}\ and\ \citenamefont {Majda}(2016)}]{Qi2016}%
  \BibitemOpen
  \bibfield  {author} {\bibinfo {author} {\bibfnamefont {D.}~\bibnamefont
  {Qi}}\ and\ \bibinfo {author} {\bibfnamefont {A.~J.}\ \bibnamefont {Majda}},\
  }\bibfield  {title} {\enquote {\bibinfo {title} {Low-dimensional
  reduced-order models for statistical response and uncertainty quantification:
  {T}wo-layer baroclinic turbulence},}\ }\href@noop {} {\bibfield  {journal}
  {\bibinfo  {journal} {J. Atoms. Sci.}\ }\textbf {\bibinfo {volume} {73}},\
  \bibinfo {pages} {4609--4639} (\bibinfo {year} {2016})}\BibitemShut {NoStop}%
\bibitem [{\citenamefont {Kaplan}\ and\ \citenamefont
  {Yorke}(1978)}]{Kaplan1978}%
  \BibitemOpen
  \bibfield  {author} {\bibinfo {author} {\bibfnamefont {J.~L.}\ \bibnamefont
  {Kaplan}}\ and\ \bibinfo {author} {\bibfnamefont {J.~A.}\ \bibnamefont
  {Yorke}},\ }\href@noop {} {\emph {\bibinfo {title} {{F}unctional
  {D}ifferential {E}quations and the {A}pproximation of {F}ixed Points}}},\
  \bibinfo {series} {{L}ecture {N}otes in {M}athem}, Vol.\ \bibinfo {volume}
  {730}\ (\bibinfo  {publisher} {Springer},\ \bibinfo {year} {1978})\ pp.\
  \bibinfo {pages} {204--227}\BibitemShut {NoStop}%
\bibitem [{\citenamefont {Lorenz}(2006)}]{Lorenz2006}%
  \BibitemOpen
  \bibfield  {author} {\bibinfo {author} {\bibfnamefont {E.~N.}\ \bibnamefont
  {Lorenz}},\ }\bibfield  {title} {\enquote {\bibinfo {title} {Predictability -
  a problem partly solved},}\ }in\ \href@noop {} {\emph {\bibinfo {booktitle}
  {Predictability of Weather and Climate}}},\ \bibinfo {editor} {edited by\
  \bibinfo {editor} {\bibfnamefont {T.}~\bibnamefont {Palmer}}\ and\ \bibinfo
  {editor} {\bibfnamefont {R.}~\bibnamefont {Hagedorn}}}\ (\bibinfo
  {publisher} {Cambridge University Press},\ \bibinfo {year} {2006})\ pp.\
  \bibinfo {pages} {40--58}\BibitemShut {NoStop}%
\bibitem [{\citenamefont {Ghia}\ \emph {et~al.}(1982)\citenamefont {Ghia},
  \citenamefont {Ghia},\ and\ \citenamefont {Shin}}]{Ghia1982}%
  \BibitemOpen
  \bibfield  {author} {\bibinfo {author} {\bibfnamefont {U.}~\bibnamefont
  {Ghia}}, \bibinfo {author} {\bibfnamefont {N.~K.}\ \bibnamefont {Ghia}}, \
  and\ \bibinfo {author} {\bibfnamefont {C.~T.}\ \bibnamefont {Shin}},\
  }\bibfield  {title} {\enquote {\bibinfo {title} {High-{R}e solutions for
  incompressible flow using the {N}avier-{S}tokes equations and a multigrid
  method},}\ }\href@noop {} {\bibfield  {journal} {\bibinfo  {journal} {J.
  Comput. Phys.}\ }\textbf {\bibinfo {volume} {48}},\ \bibinfo {pages}
  {387--411} (\bibinfo {year} {1982})}\BibitemShut {NoStop}%
\bibitem [{\citenamefont {Schreiber}\ and\ \citenamefont
  {Keller}(1983)}]{Schreiber1983}%
  \BibitemOpen
  \bibfield  {author} {\bibinfo {author} {\bibfnamefont {H.~B.}\ \bibnamefont
  {Schreiber}}\ and\ \bibinfo {author} {\bibfnamefont {H.~B.}\ \bibnamefont
  {Keller}},\ }\bibfield  {title} {\enquote {\bibinfo {title} {Driven cavity
  flows by efficient numerical techniques},}\ }\href@noop {} {\bibfield
  {journal} {\bibinfo  {journal} {J. Comput. Phys.}\ }\textbf {\bibinfo
  {volume} {49}},\ \bibinfo {pages} {310--333} (\bibinfo {year}
  {1983})}\BibitemShut {NoStop}%
\bibitem [{\citenamefont {Sahin}\ and\ \citenamefont
  {Owens}(2003)}]{Sahin2003}%
  \BibitemOpen
  \bibfield  {author} {\bibinfo {author} {\bibfnamefont {M.}~\bibnamefont
  {Sahin}}\ and\ \bibinfo {author} {\bibfnamefont {R.~G.}\ \bibnamefont
  {Owens}},\ }\bibfield  {title} {\enquote {\bibinfo {title} {A novel fully
  implicit finite volume method applied to the lid-driven cavity problem—part
  {I}: {H}igh {R}eynolds number fow calculations},}\ }\href@noop {} {\bibfield
  {journal} {\bibinfo  {journal} {J. Numer. Meth. Fluids}\ }\textbf {\bibinfo
  {volume} {42}},\ \bibinfo {pages} {57--77} (\bibinfo {year}
  {2003})}\BibitemShut {NoStop}%
\bibitem [{\citenamefont {Otto}\ and\ \citenamefont {Rowley}(2019)}]{Otto2019}%
  \BibitemOpen
  \bibfield  {author} {\bibinfo {author} {\bibfnamefont {S.~E.}\ \bibnamefont
  {Otto}}\ and\ \bibinfo {author} {\bibfnamefont {C.~W}\ \bibnamefont
  {Rowley}},\ }\bibfield  {title} {\enquote {\bibinfo {title} {Linearly
  recurrent autoencoder networks for learning dynamics},}\ }\href@noop {}
  {\bibfield  {journal} {\bibinfo  {journal} {SIAM J. Sci. Comput.}\ }\textbf
  {\bibinfo {volume} {18}},\ \bibinfo {pages} {558--593} (\bibinfo {year}
  {2019})}\BibitemShut {NoStop}%
\bibitem [{\citenamefont {Brunton}\ and\ \citenamefont
  {Kutz}(2019)}]{Brunton2019}%
  \BibitemOpen
  \bibfield  {author} {\bibinfo {author} {\bibfnamefont {S.~L.}\ \bibnamefont
  {Brunton}}\ and\ \bibinfo {author} {\bibfnamefont {J.~N.}\ \bibnamefont
  {Kutz}},\ }\bibfield  {title} {\enquote {\bibinfo {title} {{S}ingular {V}alue
  {D}ecomposition (svd) and {P}rincipal {C}omponent {A}nalysis ({PCA})},}\ }in\
  \href@noop {} {\emph {\bibinfo {booktitle} {Data-driven Science and
  Engineering: Machine Learning, Dynamical Systems, and Control}}}\ (\bibinfo
  {publisher} {Cambridge University Press},\ \bibinfo {year} {2019})\
  Chap.~\bibinfo {chapter} {1}\BibitemShut {NoStop}%
\end{thebibliography}%

\end{document}